\documentclass[tightenlines,floats,floatfix,aps,notitlepage,nofootinbib,superscriptaddress,12pt]{revtex4-1}

\usepackage[utf8]{inputenc}
\usepackage[table,dvipsnames]{xcolor}
\usepackage{multirow}
\usepackage{hyperref}
\usepackage{graphicx}
\usepackage{mathtools}
\usepackage{amsmath,amssymb,amsfonts,amsthm,latexsym,stmaryrd,physics,mathrsfs}
\allowdisplaybreaks[4]
\usepackage{marginnote}
\usepackage{color}
\usepackage{bm}
\usepackage{svg}

\usepackage{float}
\usepackage{nicefrac}
\usepackage{microtype} 
\usepackage{booktabs}
\usepackage{verbatim}
\usepackage{setspace}
\usepackage[T1]{fontenc}
\usepackage[utf8]{inputenc}
\usepackage{stfloats}
\usepackage{diagbox}
\usepackage{dsfont}
\usepackage[colorinlistoftodos,prependcaption,textsize=tiny]{todonotes}
\usepackage{epstopdf}
\usepackage{latexsym}
\usepackage{xparse}
\usepackage{caption}
\usepackage{subcaption}

\usepackage{xargs}                      %
\usepackage[colorinlistoftodos,prependcaption,textsize=tiny]{todonotes}

 \newcommand{\bq}{\begin{equation}}
 \newcommand{\eq}{\end{equation}}
 \newcommand{\bqn}{\begin{eqnarray}}
 \newcommand{\eqn}{\end{eqnarray}}

\NewDocumentCommand{\evalat}{sO{\big}mm}{%
  \IfBooleanTF{#1}
   {\mleft. #3 \mright|_{#4}}
   {#3#2|_{#4}}%
}

\def\be{\begin{eqnarray}}
\def\ee{\end{eqnarray}}

\newcommand{\ce}{\mathcal E}
\newcommand{\cf}{\mathcal F}

\newcommand{\lt}{\left}
\newcommand{\rt}{\right}

\renewcommand{\a}{\alpha}
\renewcommand{\b}{\beta}

\renewcommand{\L }{\Lambda}

\newcommand{\rmd}{\mathrm d}

\newcommand{\ltbf}{\Xi}

\begin{document}

\title{Generalized analysis of a dust collapse in effective loop quantum gravity: fate of shocks and covariance}
\author{Kristina Giesel}
\email{kristina.giesel@gravity.fau.de} 
\affiliation{Department Physik, Institut f\"ur Quantengravitation, Theoretische Physik III, Friedrich-Alexander Universit\"at Erlangen-N\"urnberg, Staudtstr. 7/B2, 91058 Erlangen, Germany}

\author{Hongguang Liu} 
\email{hongguang.liu@gravity.fau.de}
\affiliation{Department Physik, Institut f\"ur Quantengravitation, Theoretische Physik III, Friedrich-Alexander Universit\"at Erlangen-N\"urnberg, Staudtstr. 7/B2, 91058 Erlangen, Germany}

\author{Parampreet Singh}
\email{psingh@lsu.edu}
\affiliation{Department of Physics and Astronomy, Louisiana State University, Baton Rouge, LA 70803, USA}

\author{Stefan Andreas Weigl} 
\email{stefan.weigl@gravity.fau.de}
\affiliation{Department Physik, Institut f\"ur Quantengravitation, Theoretische Physik III, Friedrich-Alexander Universit\"at Erlangen-N\"urnberg, Staudtstr. 7/B2, 91058 Erlangen, Germany}

\begin{abstract}
Based on modifications inspired from loop quantum gravity (LQG), spherically symmetric models have recently been explored to understand the resolution of classical singularities and the fate of the spacetime beyond. While such phenomenological studies have provided useful insights, questions remain on whether such models exhibit some of the desired properties such as consistent LTB conditions, covariance and compatibility with the improved dynamics of loop quantum cosmology in the cosmological and LTB sectors. We provide a systematic procedure to construct effective spherically symmetric models encoding LQG modifications as a $1+1$ field theory models encoding these properties following the analysis in our companion paper \cite{LTBStabiliyPap1}. As 
concrete examples of our generalized  strategy we obtain and compare with different phenomenological models  which have been investigated recently and demonstrate resolution of singularity by quantum geometry effects via a bounce. These include models with areal gauge fixing, a polymerized vacuum solution, polymerized junction conditions and an Oppenheimer-Snyder dust collapse model. An important insight from our approach is that the dynamical equations care about the $\det(e)$ part rather than the square root of the determinant of the spatial metric. As a result, shock solutions which have been argued to exist in some models are found to be absent even if one considers coordinate transformations.
\end{abstract}

\maketitle
\noindent

\section{Introduction}
\label{sec:Intro}

\noindent
The generalization of Lema\^{\i}tre-Tolman-Bondi (LTB) models in the context of effective spherically symmetric models motivated from LQG has been a topic of active interest in the literature. In particular, various studies have been carried out formulating  models that can describe the dust collapse with quantum geometric modifications  \cite{Bojowald:2008ja,Bojowald:2009ih,Bambi:2013caa,Kelly:2020lec, BenAchour:2020bdt, Munch:2020czs,Munch:2021oqn,Li:2021snn,Husain:2021ojz,Husain:2022gwp, Munch:2022teq,Giesel:2022rxi,Bobula:2023kbo,Fazzini:2023scu}. The classical singularity in various black hole models is resolved by a bounce due to LQG effects (see for eg.  \cite{Ashtekar:2005qt,Modesto:2005zm,Boehmer:2007ket,Chiou:2012pg,Gambini:2013hna,Corichi:2015xia,Dadhich:2015ora,Olmedo:2017lvt,Ashtekar:2018lag,Han:2020uhb,Han:2022rsx,Ashtekar:2023cod}) and an interesting matter is under what assumptions some of the above models involve shock solutions \cite{Munch:2020czs, Husain:2021ojz,Husain:2011tk,Fazzini:2023scu}. In several models one considers the LTB sector already at the classical level and then quantizes this sector, see for instance \cite{Kiefer:2019csi,Kiefer:2023zxt} and in the context of LQG inspired quantizations one considers the corresponding effective model by incorporating quantum geometric modifications in a phenomenological way -- via holonomy modifications and inverse triad corrections \cite{Kelly:2020lec,BenAchour:2020bdt, Munch:2020czs,Giesel:2021dug,Li:2021snn,Husain:2021ojz,Husain:2022gwp, Han:2020uhb,Han:2022rsx, Giesel:2022rxi,Bobula:2023kbo,Fazzini:2023scu}. Following the ideas in the seminal work in \cite{Bojowald:2008ja,Bojowald:2009ih}, one can start with an effective spherically symmetric model including dust and then restrict to the (generalized) LTB sector by requiring a so-called LTB condition to hold. The results in our companion paper  \cite{LTBStabiliyPap1} have generalized the analysis in \cite{Bojowald:2008ja,Bojowald:2009ih} in several aspects.  
An important requirement of our analysis is that the LTB condition should be consistent with the effective dynamics and given this we denote such LTB conditions as compatible ones.\footnote{For details, see Definition 1 in \cite{LTBStabiliyPap1}.} In this manuscript we will show that using consistent LTB conditions in presence of different types of quantum geometric modifications lead to important insights on the fate of the spacetime in the Planck regime and the spacetime beyond the singularity. \\

\noindent
In general, compatible LTB conditions differ from the classical one by reflecting the effect of the polymerization included in terms of holonomy and inverse triad corrections. In \cite{LTBStabiliyPap1} we investigated different classes of effective models and their corresponding compatible LTB conditions. Here we want to further analyze a specific effective model for the dust collapse that falls into the class of models 
that can have holonomy as well as inverse triad corrections and has in addition the property that the effective geometric contribution to the Hamiltonian constraint and the spatial diffeomorphism constraint form a closed algebra \cite{LTBStabiliyPap1}. Furthermore, if we slightly restrict the form of the compatible LTB condition, then the effective dynamics of these models completely decouple along the radial coordinate and thus can be understood as a set of infinitely many decoupled shells each of which can be described as a Loop Quantum Cosmology (LQC) model, one for each radial coordinate, see for instance \cite{Ashtekar:2011ni} for a review on LQC models.\footnote{Models with above properties are described in Lemma 1 and Corollary 4 of our companion paper \cite{LTBStabiliyPap1}.} Based on the results in \cite{LTBStabiliyPap1}, this allows us to take a new perspective on models for effective quantum black holes and dust collapses. We can start from a given decoupled effective LQC model and construct the corresponding effective dust collapse model from it as well as the underlying spherically symmetric model which after implementing the LTB conditions yields exactly the given dust collapse model. This means, as far as the LTB condition is concerned, we can go back from the LTB model where the LTB condition is implemented to the underlying spherically symmetric model. In the case of non-marginally bound models, where the LTB condition can be understood as a gauge fixing condition \cite{LTBStabiliyPap1}, we can go from the gauge fixed model to the corresponding gauge unfixed model. Moreover, this way of constructing effective models leads to a complete model in contrast to those models where a junction condition necessarily needs to be included as for instance in \cite{BenAchour:2020bdt, Husain:2021ojz,Husain:2022gwp,Fazzini:2023scu}. Along these lines we will consider the model in \cite{Tibrewala:2012xb}, which is an example of models covered by Lemma 1 and Corollary 4 in \cite{LTBStabiliyPap1}. Here in the LTB sector we use the decoupled improved dynamics of LQC as our starting point and construct the underlying spherically symmetric model and dust collapse model which we show to  
agree with the model in \cite{Husain:2021ojz,Husain:2022gwp} when we apply an areal gauge.\\

\noindent
To know the underlying effective spherically symmetric model has several advantages that yield new insights on the properties and  relation to other models available in the literature. First, we can link that model to its corresponding extended mimetic model \cite{Chamseddine:2013kea,Sebastiani:2016ras,Takahashi:2017pje,Langlois:2017mxy}. Using the results from \cite{BenAchour:2017ivq,Han:2022rsx} we can provide a covariant Lagrangian formulation of the effective model. Second, with the knowledge of the covariant model and its relation to the spherically symmetric model we can perform coordinate transformations involving the temporal coordinate. As is usual in mimetic gravity models, a temporal gauge fixing is implemented with respect to the involved scalar field, that apart from gravity is one of the elementary degrees of freedom. We can understand these kinds of coordinate transformations as being encoded in the corresponding transformations of the scalar field. This allows us to discuss the effective dynamics of that model in the generalized Gullstrand–Painlevé, generalized Schwarzschild-like coordinates as well as LTB coordinates and discuss new insights we obtain by considering different sets of coordinates. 
~\\
~\\
As in the model of \cite{Husain:2021ojz,Husain:2022gwp}, in this paper we concentrate on the marginally bound case. Investigating the model from the covariant perspective as well as from different sets of coordinates will obtain new insights on the presence of shock solutions in the model \cite{Husain:2021ojz,Husain:2022gwp}. We find that demanding that the model agrees with the improved dynamics of LQC \cite{Ashtekar:2006wn} in the cosmological sector and on each decoupled shell with consistent LTB conditions results in the Hamiltonian constraint with a $\mathrm{det}\, e$ term. This is in contrast to the terms containing square roots of the spatial metric in \cite{Husain:2021ojz,Husain:2022gwp}. As a result, shock solutions are absent even if one works solely in the Gullstrand–Painlevé coordinates. While this work was in the final stages of preparation, a work exploring the presence of shock solutions has appeared which attributes it as a coordinate artifact \cite{Fazzini:2023scu}. Our results show that above is the more fundamental reason why shock solutions do 
 not exist independent of the coordinate system. \\

\noindent
A detailed analysis of the effective dynamics allows us to obtain an analytical solution in LTB coordinates that can also encode inhomogeneous dust and thus extends the solution presented in \cite{Giesel:2021dug,Fazzini:2023scu}. We discuss the resulting polymerized vacuum solution and show the fact that the curvature invariants are non-vanishing in contrast to the vacuum solution in the classical theory. This can be understood by the properties of the underlying mimetic model which corresponds to this effective model. Furthermore we show that by applying an appropriate coordinate transformation to the polymerized vacuum solution presented here, we can rediscover the form of the metric of the models in \cite{Kelly:2020uwj,Lewandowski:2022zce}. Thus the model in this work provides a consistent $1+1$--dimensional model beyond the junction conditions used in \cite{Lewandowski:2022zce}. Another novel aspect of this work is that to the knowledge of the authors the existing models in the literature so far still use classical junction conditions or weak solutions to describe quantum dust collapses. Here we take advantage of the knowledge about the underlying mimetic model and its modified Einstein's equation to derive the closed formula of polymerized junction condition in the effective theory. Compared to the classical models such effective junction conditions differ from their classical counterparts by additional contributions coming from the polymerization. This opens up the possibility to discuss the junction conditions used in \cite{Fazzini:2023scu,Husain:2022gwp} in the context of the quantum Oppenheimer-Snyder dust collapse and show that these choices violate the smoothness condition of the (additional) scalar field involved in the mimetic model which plays the role of the temporal reference field in the effective model.\\

\noindent
The paper is structured as follows. In section \ref{sec:EffModfromCosm}  we discuss how the underlying effective spherically symmetric and LTB model  can be obtained from a given choice of uncoupled effective LQC models. Afterwards in subsection \ref{sec:CovLagrangian} we present the corresponding covariant Lagrangian of the mimetic model of the effective model derived in \ref{sec:EffModfromCosm}. A detailed investigation of the effective dynamics of the model derived in section \ref{sec:EffModfromCosm} is presented in section \ref{sec:EffDyn}. In subsection \ref{sec:TrappSurf} we discuss under which conditions trapped surfaces form. The polymerized vacuum solution and its physical properties is the topic of subsection \ref{sec:PolyVacuum}. Subsection \ref{sec:PolyJuncCond} contains the derivation of the polymerized junction condition using the modified Einstein's equations of the underlying mimetic model.  This junction condition is then applied to the effective model describing an Oppenheimer-Snyder collapse in subsection \ref{sec:EffOppSny}. In section \ref{sec:Comparison} we use the results from section \ref{sec:EffDyn} and compare those with existing models in the literature in several aspects. In subsection \ref{sec:ArialGauge} we show that the resulting model obtained in \ref{sec:EffModel} can be understood as  -- with respect to the areal gauge -- a gauge unfixed version of the model in  \cite{Husain:2021ojz,Husain:2022gwp}. We present a comparison of the existing different polymerized vacuum solutions in subsection \ref{sec:CompPolyVac} and can link different models by applying a suitable coordinate transformation by means of the knowledge of the underlying covariant extended mimetic model. 
As existing models consider classical junction conditions or weak solutions, we compare the different existing junction conditions in the context of the effective Oppenheimer-Snyder dust collapse with the polymerized junction condition obtained in this work and discuss their properties in subsection \ref{sec:CompOSJuncCond}. Finally we give a summary and outlook in section \ref{sec:Concl}.

\section{Effective spherically symmetric model from links to cosmological dynamics}
\label{sec:EffModfromCosm}
\noindent
This section lays the foundations for the forthcoming analysis of the dynamics and resulting structures of spacetimes with quantum gravitational modifications. We will briefly review spherically symmetric spacetimes with dust and LTB spacetimes in the context of GR. Based on the results of our companion paper \cite{LTBStabiliyPap1} we will then derive in subsection \ref{sec:EffModel} starting from a choice of along the radial coordinate decoupled LQC models, the corresponding effective spherical symmetric model which after implementing the classical LTB condition merges dynamically into the given LQC models. In subsection \ref{sec:ArialGauge} we show that by implementing the areal gauge into the previously derived effective spherical symmetric model we obtain the dust collapse model analyzed in \cite{Husain:2021ojz,Husain:2022gwp}. Finally we present in subsection \ref{sec:CovLagrangian} an underlying covariant Lagrangian of the previously derived effective model in the class of extended mimetic models.\\

\noindent
In LQG we work with Ashtekar-Barbero variables and in the case of spherical symmetry we can choose the spatial manifold to be $\mathbb{R}\times S^2$ and write the Ashtekar-Barbero connection $A_a^j$ as well as the densitized triad variables $E^j_a$ after implementing the Gau{\ss} constraint following \cite{Ashtekar:2005qt,Modesto:2005zm,Boehmer:2007ket,Chiou:2012pg,Gambini:2013hna,Corichi:2015xia,Dadhich:2015ora,Olmedo:2017lvt,Ashtekar:2018lag,Han:2020uhb,Ashtekar:2023cod}
as
\begin{align}
A_a^j \tau_j \mathrm{~d} X^a & =  2\beta K_x(x) \tau_1 \mathrm{~d} x+\left(\beta K_\phi(x) \tau_2+\frac{\partial_x E^x(x)}{2E^\phi(x)} \tau_3\right) \mathrm{d} \theta \\
&\quad+\left(\beta K_\phi(x) \tau_3-\frac{\partial_x E^x(x)}{2E^\phi(x)} \tau_2\right) \sin (\theta) \mathrm{d} \phi+\cos (\theta) \tau_1 \mathrm{~d} \phi, \nonumber\\
    E_j^a \tau^j \frac{\partial}{\partial X^a} & = E^x(x) \sin (\theta) \tau_1 \partial_x+\left(E^\phi(x) \tau_2\right) \sin (\theta) \partial_\theta+\left(E^\phi(x) \tau_3\right) \partial_{\phi}\,, 
\end{align}
where $\beta$ is the Barbero-Immirzi parameter and $\tau_j=-\frac{1}{2}\sigma_j$ with $\sigma_j$ being the Pauli matrices as well as $X^a=(x,\theta,\phi)$ with $a=1,2,3$ denoting the spherical coordinates. The triad is given by
\begin{eqnarray}
     e_a^j \tau_j dX^a & = \frac{E^{\phi}(x)}{\sqrt{E^x(x)}} \tau_1 d x+\left( \sqrt{E^x(x)}\tau_2\right)  d \theta+\left( \sqrt{E^x}\tau_3\right) \sin(\theta) d {\phi}\, ,
\end{eqnarray}
such that we have $\det e = E^{\phi}(x)\sqrt{E^x(x)} \sin(\theta)$. We will show later on in section \ref{sec:EffModel} and \ref{sec:PolyVacuum} such choice is favored by the requirement that the effective dynamics agrees with the improved dynamics of LQC on each decoupled shell in LTB coordinate and we have smooth trajectories in phase space.
The phase space is defined by the the following Poisson algebra
\begin{align}\label{eq:cano_spherical}
    \{K_x(x), E^x(y)\} = G \delta(x,y), \quad \quad \{K_\phi(x), E^\phi(y)\} =G \delta(x,y),
\end{align}
where $G$ is Newton's constant. In addition to the gravitational degrees of freedom we consider a dust field $T$ together with is conjugate momentum $(T,P_T)$ that will serve as a temporal reference field in the classical model. The dust model we consider is non-rotational dust which can can understood as a special case of the Brown Kucha\v{r} dust model \cite{Kuchar:1995xn}. The details on the dust model are not important for this work and for the interested reader they can be found in section 2 in \cite{LTBStabiliyPap1}. In these variables the general spherically symmetric metric has the form of a generalized Gullstrand–Painlevé metric \cite{Gullstrand:1922tfa}\cite{painleve1921mecanique}:
\begin{eqnarray}
 \label{eq:metric}
\rmd s^2=-\rmd t^2+\L(t,x)^2\left( \rmd x^2 + N^x(t,x) dt^2 \right)+R(t,x)^2\lt[\rmd\theta^2+\sin^2(\theta)\rmd\varphi^2\rt],  
\end{eqnarray}
with $\L(t,x)\coloneqq\frac{E^\varphi}{\sqrt{|E^x|}}$ and $R(t,x)\coloneqq\sqrt{|E^x|}$\,. A special class spherical symmetric solutions is Einstein gravity coupled to dust. These admit  the so-called LTB condition
\be \label{eq:LTBcondition}
E^{\phi} = \frac{ \partial_x E^x }{ 2 \ltbf } =  \frac{ \partial_x E^x }{ 2\sqrt{1+\mathcal{E}(x)} }\,, 
\ee 
where the LTB function used in \cite{LTBStabiliyPap1} is related to $\mathcal{E}$ by $\ltbf(x):= \sqrt{1+\mathcal{E}(x)}\,$. The function $\mathcal{E}(x) > -1$ is a measure of the total energy of the dust shell at radial coordinate $x$ (see for example the discussion in \cite{Szekeres:1995gy} or \cite{Lasky:2006hq}). When we have $\mathcal{E}(x_0)=0$, then such a shell has zero kinematical energy at infinity, i.e. it is at rest. One can also see this in the dynamical equation of $R$ in \eqref{eq:metricLTBchoice} for the limit of the physical radius ${R(t,x)}\rightarrow \infty$. For positive or negative $\mathcal{E}(x_0)$ the kinematical energy is higher or lower than the negative gravitational potential. Depending on the sign choice of the square root (which is related to the direction of time) the shells can either expand or collapse.\\

\noindent
The metric corresponding to \eqref{eq:LTBcondition} is given by the LTB spacetime \cite{lematre,tolman,bondi} which is usually written in the form of \eqref{eq:metric} with \cite{Giesel:2009jp}
\begin{eqnarray}\label{eq:metricLTBchoice}N^x(t,x) = 0 , \quad \Lambda(t,x)^2=\frac{\lt[\partial_xR(t,x)\rt]^2}{1+\ce(x)},\quad \partial_t R(t,x)=\pm\sqrt{\ce(x)+\frac{\cf(x)}{R(t,x)}}\,.
\end{eqnarray}
The functions $\ce(x),\cf(x)$ are arbitrary and have to be adapted to the dust profile one wants to consider. Note that the case $\partial_xR(t,x)=0$ is known as a shell crossing singularity. Working with positive dust densities these singularities can be completely avoided by choosing appropriate dust profiles \cite{Hellaby:1985zz}. The interpretation of $\ce(x)$ was already discussed beforehand and the one of $\cf(x)>0$ is that it is the total gravitational mass within the sphere at radial coordinate $x$. The solution of the partial differential equation in \eqref{eq:metricLTBchoice} can be parameterized by the different regimes of $\ce(x)$:
\begin{eqnarray}\ce(x)>0:&&\quad R\left(t, x\right) =\frac{\cf\left(x\right)}{2 \ce\left(x\right)}\lt[\cosh (\eta)-1\rt], \quad\sinh (\eta)-\eta =\frac{2\left[\ce\left(x\right)\right]^{\frac{3}{2}}\left(\tilde\beta\left(x\right)-t\right)}
{\cf\left(x\right)},\label{positiveE}\\
\ce(x)=0:&&\quad R\left(t, x\right)=\left[\frac{3}{2} \sqrt{\cf(x)}\left(\tilde\beta\left(x\right)-t\right)\right]^{2 / 3},\label{zeroE}\\
\ce(x)<0:&&\quad R\left(\tau, x\right) =\frac{\cf\left(x\right)}{2|\ce\left(x\right)|}\lt[1-\cos (\eta)\rt], \quad \eta-\sin (\eta) =\frac{2|\ce\left(x\right)|^{\frac{3}{2}}\left(\tilde\beta\left(x\right)-t\right)}{\cf\left(x\right)}\,.\label{negativeE}
\end{eqnarray}
In the arbitrary function $\tilde \b(x)$ is the residual gauge freedom of choosing different sets of radial coordinates encoded. Note that at the points $t=\tilde \b(x)$ the metric is singular. In the marginal bound case, i.e. choosing $\ce(x)\equiv0$, when further considering the constant mass distribution $\cf=R_s=2GM$ and using the gauge $\tilde \b(x)=x$, we regain the Schwarzschild metric in Lema\^{\i}tre coordinates.

\subsection{Derivation of the underlying effective spherically symmetric model}
\label{sec:EffModel}
\noindent 
As discussed in the introduction, the results in our companion paper \cite{LTBStabiliyPap1} allow us to take an effective polymerized cosmology model, e.g. effective LQC model, as a starting point and construct the underlying spherically symmetric model that we assume to be, as a spherically symmetric model, decoupled along the radial coordinate. This spherically symmetric model has the property that when we implement the LTB condition that restricts the model to its effective LTB sector, we obtain for each radial coordinate the decoupled LQC model we took as a starting point.  Hence, we can characterize such effective spherically symmetric models by its decoupled LTB model. This class of models is the one described in Lemma 1 and Corollary 4 in our companion paper \cite{LTBStabiliyPap1}.
A concrete example of the model that we closely investigate is the model that was briefly discussed as the second example in section 4 of \cite{LTBStabiliyPap1}. Defining the elementary variables $b$ and $v$ as
\be \label{eq:defbandv}
b(t,x) = \frac{K_{\phi}}{\sqrt{E^x}}(t,x), \quad v(t,x) =(E^x)^{3/2}(t,x) \,,
\ee 
the effective dynamical equations of the model in LTB sector are given by 
\begin{eqnarray}
\label{reduced equations of motion}
\partial_t v &=& \frac{3 v \sin \left(2 \a  b \right)}{2 \a },\quad  \partial_t b =- \frac{1}{2} \left(\frac{\mathcal{E}(x)}{v^{\frac{2}{3}}}+\frac{3 \sin ^2\left(\a b\right)}{\a^2}\right)  \,.
\end{eqnarray}
Since there are no spatial derivatives appearing,  this set of differential equations is completely decoupled along the radial coordinate. For every choice of an LTB function $\mathcal{E}(x)$ we can solve these equations for each $x=x_0$ independently. In the context of dust models this means that every shell can be treated independently of the others, there is no compression or crossing of shells.\\

\noindent
Based on the results in \cite{LTBStabiliyPap1} we want to now construct a spherically symmetric effective model that has exactly above stated equations of motion in the LTB sector. For this purpose we need to find the corresponding effective gravitational part of the Hamiltonian constraint denoted by $C^\Delta$ according to equation (3.3) and (3.6) in \cite{LTBStabiliyPap1} and then  the corresponding partially gauge fixed primary Hamiltonian in \cite{LTBStabiliyPap1}. The latter is the canonical Hamiltonian of the effective spherically symmetric model after implementing the comoving gauge with respect to the dust. The total Hamiltonian constraint before the temporal gauge fixing is given by $C^\Delta$ and the contribution of the dust and likewise for the spatial diffeomorphism constraint that before the gauge fixing is applied also has a non-vanishing contribution from the dust.
~\\

\noindent
One can see from these equations on one hand we need to find an appropriate polymerization function ${f}$ that includes holonomy and in general inverse triad corrections in those terms that involve extrinsic curvature variables. On the other hand we also need to know the inverse triad corrections encoded  in  $h_1$ and $h_2$, that are involved in the remaining terms that do not depend on the extrinsic curvature.
As mentioned above we will restrict ourselves to effective models that fall in the scope of Lemma 1 in \cite{LTBStabiliyPap1}, which means that $C^\Delta$ is a conserved quantity. This puts additional constraints on the polymerization functions ${f}$ and inverse triad corrections, see eq. (3.12) in \cite{LTBStabiliyPap1}. Consequently, $f$ can be split into two separated parts ${f}^{(1)}, {f}^{(2)}$ which do not encode $K_x$ polymerizations anymore. According to equation (3.11) and (3.13) in \cite{LTBStabiliyPap1} we can now write $C^\Delta$ as 
\begin{eqnarray}
   C^{\Delta}= - \frac{E^{\phi}\sqrt{E^x}}{2 G }\left[ \frac{{f}^{(1)}(K_{\phi},E^x)}{E^x} + \frac{4 K_x {f}^{(2)}(K_{\phi},E^x)}{E^{\phi}}+h_1 \frac{1 -\qty(\frac{  {{E^x}}'}{2{{E^{\phi}}} })^2}{E^x}  - \frac{2 h_2}{E^\phi}\Big(\frac{  {E^x}'}{2{{E^{\phi}}} }\Big)'\right]\,.
\end{eqnarray}
Then with the help of Corollary 4 of \cite{LTBStabiliyPap1} we can directly read off ${f}^{(1)}$ from the differential equation of $E^x$ (respectively $v$). Excluding the presence of inverse triad corrections, i.e. $h_1 = h_2 =1$, we can then determine ${f}^{(2)}$ and the corresponding compatible LTB condition.  It turns out that the effective compatible LTB condition is exactly the classical one, i.e. $g_\Delta(x) = \ltbf(x) = \sqrt{1+\mathcal{E}(x)}$ and the remaining two polymerization functions have the form
\begin{eqnarray}
  && {f}^{(1)} = \frac{3 E^x \sin\left(\frac{\alpha K_{\phi}}{\sqrt{E^x}}\right)^2}{\alpha^2 } - 2 K^\phi f^{(2)} \ , \quad {f}^{(2)} = \frac{\sqrt{E^x}\sin\left( \frac{2 \alpha K_{\phi}}{\sqrt{E^x}}\right)}{2 \alpha}\,,\quad \alpha := \beta\sqrt{\Delta}
\end{eqnarray} 
with $\Delta$  being the minimal area gap in  LQC \cite{Ashtekar:2011ni}.
~\\
Putting these results back in the partially gauge fixed primary Hamiltonian we end up with
\begin{eqnarray}
\label{eq:effectiveHamiltonian}
\mathcal{H}^{\Delta}_{pri} &=& \int \dd{x} \,\qty( {C}^{\Delta} + N^x C_x) \\
   C^{\Delta} &=&- \frac{E^{\phi} \sqrt{E^x}}{2G} \bigg[\frac{3}{\a^2} \sin ^2\left(\frac{\a K_{\phi}}{\sqrt{{E^x}}}\right) +\frac{ (2 {E^x} K_x-{E^{\phi}} K_{\phi})}{\a \sqrt{{E^x}} {E^{\phi}}}  \sin \left(\frac{2 \a K_{\phi}}{\sqrt{{E^x}}}\right)+\\
   &&\quad\quad\quad\quad\quad+ \frac{1 -\qty(\frac{  {{E^x}}'}{2{{E^{\phi}}} })^2}{E^x}  - \frac{2}{E^\phi}\Big(\frac{  {E^x}'}{2{{E^{\phi}}} }\Big)'\bigg] \nonumber\,.
\end{eqnarray}
The class of models we consider have the property that the diffeomorphism constraint $C_x$ is not polymerized and therefore takes its classical form
\begin{eqnarray}\label{eq:diffeo}
    C_x = \frac{1}{G} ( E^{\phi} K_{\phi}' -K_x (E^x)').
\end{eqnarray}
Note that we have not yet implemented the compatible LTB condition. One can check that this polymerized geometric contribution to the Hamiltonian constraint $C^{\Delta}$ agrees with the one presented in \cite{Tibrewala:2012xb}. It is straightforward to verify that the model is invariant under gauge transformations generated by the classical diffeomorphism constraint $C_x$ along the $x$-direction. As the model falls into the class of models governed by Lemma 1, meaning $C^{\Delta}$ is a conserved quantity, we explicitly have
\be
&&\{C^{\Delta}[N_1], C^{\Delta}[N_2] \} = \qty( \cos(\frac{2 \a K_{\phi}}{\sqrt{{E^x}}}) \frac{E^x }{(E^{\phi})^2} C_x) [N_1 N_2' -N_1' N_2]\,. 
\ee 
This means the algebra of $\big(C^{\Delta},C_x\big)$ is a closed algebra. Compared to the classical case, where for a vanishing shift vector we can identify $C^\Delta$ with the physical Hamiltonian density of the temporally gauge fixed model,  two Hamiltonian densities just yield the spatial diffeomorphism with an appropriate smearing function whereas the effective model includes an additional deformation factor $\cos(\frac{2 \a K_{\phi}}{\sqrt{{E^x}}})$.\\

\noindent
To conclude this section we discuss the corresponding effective LTB model. Based on the results in Corollary 4 in \cite{LTBStabiliyPap1} and using the classical LTB condition, in the notation of the corollary this means $\tilde{g}_{\Delta} = 1$, as a compatible LTB condition\footnote{This means in particular that the compatible LTB condition is conserved with respect to the dynamics produced by the effective primary Hamiltonian given in eq. \eqref{eq:effectiveHamiltonian}.}, we can rewrite $C^{\Delta}$ as
\be \label{eq:CdeltaeffectiveLTBmodel}
C^{\Delta} = -\frac{1}{2G} \left(\frac{\mathcal{E}'(x)}{2 (1+ \mathcal{E}(x))} + \partial_x \right) \widetilde{C}^{\Delta} \ , 
\ee
with
\be
\widetilde{C}^{\Delta} = \frac{\sqrt{E^x}}{\sqrt{1+\mathcal{E}(x)}}\left( \frac{E^x}{\alpha^2} \sin^2 \left(\frac{\a K_{\phi}}{\sqrt{E^x}}\right) - \mathcal{E}(x) \right)\,. 
\ee 
In the non-marginally bound case, the dynamics can be computed with the corresponding Dirac bracket where we reduce with respect to the classical diffeomorphism constraint $C_x$ and the LTB gauge fixing condition $G_x^{\Delta}(x)= \frac{E^x{}'}{2 E^{\phi}}(x) -\sqrt{1+\mathcal{E}(x)}$. The result is
\begin{equation}
    \poissonbracket{K_{\phi}(x)}{E^x(y)}_D = - 4 G \frac{ (1+ \mathcal{E}(x))^{\frac{3}{2}} }{\mathcal{E}'(x)} \delta(x-y)\,.
\end{equation}
For the marginally bound case we set $\Xi=1$ in $C^{\Delta}$ and  use the usual Poisson bracket to derive the resulting equations of motion.

\subsection{Covariant Lagrangian of the effective theory}
\label{sec:CovLagrangian}
\noindent
As shown in \cite{BenAchour:2017ivq,Han:2022rsx}, the effective Hamiltonian presented in the previous sections can be generated from a covariant Lagrangian in the class of extended mimetic theories \cite{Chamseddine:2013kea,Sebastiani:2016ras,Takahashi:2017pje,Langlois:2017mxy}. In the spherically symmetric case, such a model can be derived from a two-dimensional (2D) action of the form
\begin{eqnarray}\label{eq:cov_action}
    S_2&=&\frac{1}{4G}\int_{\mathscr{M}_2} \rmd^2 x\, \det(e) \big\{ e^{2\psi}\left(R_{h}+2h^{ij}\partial_{i}\psi\partial_{j}\psi\right)+2 
    +\, e^{2\psi}\big[ L_{\phi}\left(X,Y\right)+\lambda\left(\partial_j \phi \partial^j \phi+1\right)\big]\big\}\,, \nonumber \\
&&
\end{eqnarray}
where $h_{ij}$ is the 2D metric, $\det(e)$ denotes the determinant of 2D triad field and $R_h$ is 2D scalar curvature. The quantities $X,Y$ consist of couplings between the metric $h_{ij}$ and derivatives of $\phi,\psi$ and are given by 
\be\label{eq:choiceofXandY}
X = - \Box_{h}\phi + Y \ , \quad  Y = - h^{ij}\partial_{i}\psi\partial_{j}\phi\ .
\ee
Note that usually one defines the action by means of the determinant of the metric $\sqrt{-h}$, which is even under parity transformation. However, here it is crucial to use $\det(e)$ instead, which is an odd function under parity transformations. It turns out that this is necessary to reproduce the correct effective dynamics of the model, because $E^{\phi}$ may change sign before and after the bounce. We want to remark that in principle we can have a mimetic coupling $\lambda\left(\partial_j \phi \partial^j \phi+1\right)$ with a different behavior under parity transformation, but this will not influence the dynamics after gauge, as $\lambda$ is just a Lagrangian multiplier. The mimetic potential encoded in the function $L_{\phi}$ reads
\begin{eqnarray}
    L_{\phi}^{(m)}(X,Y) = \frac{3}{2 \alpha^2} + 2 XY -Y^2 -\frac{3\,{}_{(m)}\sqrt{1-4 \alpha^2 Y^2}}{2 \alpha^2} - (X+Y) \frac{\sin^{-1}_{(m)}(2 \alpha Y)}{\alpha}, \quad m \in \mathbb{Z}\,.
\end{eqnarray}
The label $(m)$ indicates that $L_{\phi}$ is defined as a multi-valued function due to the involved $\sin^{-1}_{(m)}(\eta)$ and ${}_{(m)}\sqrt{\eta}$ which are respectively defined as
\be\label{eq:defmultivaluesin}
\sin^{-1}_{(m)}(\eta)&=&(-1)^m\arcsin(\eta)+m\pi\,\,\in \lt[m\pi-\frac{\pi}{2},m\pi+\frac{\pi}{2}\rt],\quad \ \eta\in[-1,1]\,, \\
{}_{(m)}\sqrt{\eta} &=& (-1)^m \sqrt{\eta} \,\,\,\, \in  (-1)^m [0, +\infty)\,.
\ee
In order to match the LTB metric in the comoving gauge, $X$ and $Y$ take the following form
\begin{eqnarray}
    X = \frac{\partial_t E^{\phi}}{ E^{\phi}}\,  , \qquad Y = \frac{ \partial_t E^{x}}{ 2 E^{x}}  = \frac{\sin( 2 \alpha b) }{2 \alpha}\,,
\end{eqnarray}
where we used the equation of motion for $v$ in \eqref{reduced equations of motion} in the second relation. 
Note that due to the periodicity of the sine function there are multiple arguments of $b$ possible for a chosen value of $Y_0$. We can also write this property as a symmetry transformation
\be
b \mapsto (-1)^m b+\frac{m\pi}{2\alpha}, \qquad m \in\mathbb{Z}
\ee
of the function $Y(b)$. Interestingly, plugging above choice of $Y$ back in the mimetic potential $L_{\phi}$ we can drop the multi-value label, since the symmetry of $b$ exactly cancels it. Finally we end up with the single-valued potential term
\be
L_{\phi}(X,Y,K_x,K_\phi) = && \frac{3}{2 \alpha^2} + 2 XY -Y^2 -\frac{3 \cos\left( 2 \alpha b \right)}{2 \alpha^2} - (X+Y) 2 b\,.
\ee
The canonical analysis in the comoving gauge with $\phi =t$ and the equivalence with the effective primary Hamiltonian in \eqref{eq:effectiveHamiltonian} is presented in Appendix \ref{app:cano_analysis}.
As will be analyzed later on section \ref{sec:MBEffMod}, at the bounce where $2 \alpha b = \pi$ we have $Y = 0$ and $m=1$, since for this label $2 \alpha b$ takes values in $[\frac{\pi}{2}, \frac{3\pi}{2})$. In this case we still deviate from classical theory in terms of the variables $(X,Y)$ as there is a non-trivial $m$.
~\\

\noindent
The equations of motion of the mimetic model are given by the pendant of Einstein's equations and the mimetic condition, which can be written as
\begin{eqnarray}\label{eq:cov_eq}
   G^{\Delta}_{\mu\nu} := G_{\mu\nu}- T^{\phi}_{\mu\nu}=-2\lambda\partial_{\mu} \phi\partial_{\nu} \phi, \quad \partial_{\mu} \phi \partial^{\mu} \phi = -1 \,.
\end{eqnarray}
The right hand side of the modified Einstein equation is the energy momentum tensor of non-rotational dust, where $2\lambda$ is the dust energy density.
One important property of this model is that the dynamics of the field $\phi$ are not independent but determined by the mimetic condition and the (generalized) Einstein's equations. As a result, performing a gauge fixing for $\phi$ commutes with the variation of the action, as shown in \cite{Han:2022rsx}. The dynamical part of $\phi$ is exactly the one of non-rotational dust, which is a special case of the Brown-Kucha\v{r} dust introduced in \cite{Kuchar:1995xn}. The temporal gauge fixed primary Hamiltonian $\mathcal{H}^{\Delta}_{pri}$ can be obtained from the model by applying the temporal gauge fixing $\phi = t$ similar to what one does in deparameterized models with one or more dust fields. However, not these are not the same physical reference fields as the coupling to the gravitational sector is different in these cases.
~\\
An advantage of the existence of the link between effective model and the general covariant mimetic Lagrangian is that it
opens up the possibility of considering coordinate transformations $t \to \tau(t,x)$ which are encoded in the transformation of the scalar field $\phi(t) \to \phi(\tau,x)$. Moreover, we note that in the comoving gauge we have
\begin{eqnarray}
    2\lambda E^{\phi} \sqrt{E^x}= C^{\Delta}
\end{eqnarray}
and thus the dust energy density $\lambda$ is just given by the value of $C^{\Delta}$ in the comoving gauge. This implies for polymerized vacuum solution given by $\tilde{C}^{\Delta} = 0$ is coordinate independent and corresponds to $\lambda =0$ thus vanishes the non-rotational dust energy.
~\\
With the knowledge of the underlying covariant mimetic model discussed in subsection \ref{sec:CovLagrangian} for the effective model introduced in subsection \ref{sec:EffModel} we are able to examine the properties of the effective dynamics from a broader perspective in the next section.
\section{Effective dynamics}
\label{sec:EffDyn}
\noindent
In this section we investigate the dynamics of the effective model constructed in section \ref{sec:EffModel}. First we take advantage of the fact that the effective spherically symmetric model when restricted to its LTB sector consists of an infinitely many decoupled effective LQC models, one for each radial coordinate $x$, which enables us to obtain an analytical solution for the marginally bound sector even in the case of an inhomogeneous dust profile. We first discuss the properties of the marginally bound solution in subsection \ref{sec:MBEffMod}. The condition under which trapped surfaces form is discussed in subsection \ref{sec:TrappSurf} both for the marginally and non-marginally bound cases. In subsection \ref{sec:PolyVacuum} we present the polymerized vacuum solution, corresponding to the vacuum solution of the underlying mimetic model and compare its properties to those of the vacuum solution of the classical theory. By means of the knowledge of the underlying mimetic model discussed in subsection \ref{sec:CovLagrangian}, we derive a polymerized junction condition for the effective model that goes beyond the classical one and we consider as one application the effective Oppenheimer-Snyder dust collapse in subsection \ref{sec:EffOppSny}. Given the results in subsection \ref{sec:MBEffMod}-\ref{sec:EffOppSny} a detailed comparison of the results in this work with existing models in the literature can be found in section \ref{sec:Comparison}.
~\\
~\\
To start with we rewrite the infinitely many decoupled equations of motion \eqref{reduced equations of motion} in LTB coordinates with $N^x = 0$, as decoupled modified Friedmann equations at each $x$ with the help of \eqref{eq:CdeltaeffectiveLTBmodel} similar as it is done in LQC
\begin{eqnarray}
\label{Friedmann from homogeneous reduction}
\frac{\dot{R}^2}{R^2}(x) =\left( \frac{\kappa  \rho }{6} + \frac{\mathcal{E}}{R^2} \right) \left( 1 - \alpha^2\left( \frac{\kappa  \rho }{6} + \frac{\mathcal{E}}{R^2} \right) \right)(x) \,,
\end{eqnarray}
where we define $\rho = \frac{4 \pi}{3 v} \widetilde{C}^{\Delta} $ and $R = v^{\frac{1}{3}}$.
Given equation \eqref{Friedmann from homogeneous reduction} we have that at each point $x$ a symmetric bounce will happen at $\dot{E}^x = 0$ with a certain critical density. As a result, this predicts a minimal area of the model which takes the value $\min(E^x|_{\dot{E}^x = 0})$.  
\subsection{Effective dynamics for the marginally bound case}
\label{sec:MBEffMod}
\noindent
The effective equation \eqref{Friedmann from homogeneous reduction} can be integrated out analytically. For the choice $\mathcal{E}(x) = 0$, i.e. the marginally bound case, this leads to the following solutions parameterized by an arbitrary function $\tilde\beta(x)$ that depends on the radial coordinate $x$ only:
\begin{eqnarray}\label{eq:general_solution}
    R(x,t) = \left( \mathcal{F}(x) \left( \frac{9}{4} (\tilde\beta(x) - t )^2 +  \alpha^2 \right) \right)^{\frac{1}{3}} 
\end{eqnarray}
 where $\mathcal{F}(x) = 2 G M(x) := 8 \pi G \widetilde{C}^{\Delta}(x)$. The choice $\alpha =0$ relates to the classical LTB solutions in the marginally bound case, see eq. \eqref{eq:metricLTBchoice}. For the special case of a homogeneous dust profile the analytical solution has been already presented in \cite{Giesel:2022rxi} and \cite{Fazzini:2023scu}\footnote{The work in \cite{Fazzini:2023scu} which appeared as this work was completed also addresses the question of shock solutions for effective dust models. We discuss in detail in section \ref{sec:Comparison} the relation to the results presented in this work.}. Choosing $\mathcal{F}$  to be constant  and $\mathcal{F} = R_s = 2 G m_s$ corresponds to the vacuum solution where $\widetilde{C}^{\Delta} = 0$, in this case we can choose $\tilde\beta(x) = x$ as this is just a rescaling of the radial coordinate $x \to \tilde\beta(x)$.
~\\
Note that equation \eqref{Friedmann from homogeneous reduction} is not sensitive to the direction of time due to the involved squares and consequently the solution \eqref{eq:general_solution} carries the same property. As a result, these are also solutions to a mimetic model similar to \eqref{eq:cov_action} but coupled with $\sqrt{-h}$, which is an even function under parity transformation, instead of $\det(e)$ (see also the discussion after eq. \eqref{eq:choiceofXandY}). However, from the Hamiltonian we started with, a sign change is possible for $E^{\phi}$. For example in case of the vacuum solution this happens at the bounce and the difference can be seen in the variable $b$ as shown in subsection \ref{sec:CompPolyVac}.We remark that from Hamilton's equation it is clear that we have
\begin{eqnarray}
    2 \alpha b = \sin^{-1}_{(m)} \left( 2 \alpha \frac{\partial_t R}{R} \right)  = \sin^{-1}_{(m)} \left(-\frac{6\alpha (\beta (x)-t)}{\alpha^2+\frac{9}{4} (t-\beta (x))^2} \right)\,.
\end{eqnarray}
For the classical solution we have $m=0$ and from \eqref{reduced equations of motion} we realize that $b$ will monotonically increase in the marginally bound case. Since we have $2 \alpha b < \pi$ before the bounce whereas $2 \alpha b > \pi$ after the bounce, equation \eqref{eq:defmultivaluesin} tells us that we need to have $m=1$ at the bounce. 
This holds for all solutions including the polymerized vacuum solution and the homogeneous dust solution. From this point of view, the fact that in \cite{Husain:2022gwp}  the $b$ function in the vacuum case stays in  the range $2 \alpha b < \pi$ is an effect of using the coordinate $r=R(t,x)$ which is not sensitive to the sign. The detailed comparison will be given in subsection \ref{sec:ArialGauge}.
~\\
The metric corresponding to the solution \eqref{eq:general_solution} reads according to \eqref{eq:metric} and \eqref{eq:metricLTBchoice} in the marginally bound case
 \begin{eqnarray}
     \mathrm{d}s^2 = - \mathrm{d}t^2 + \left( {\partial_x R} \right)^2 \mathrm{d}x^2 + R^2 \mathrm{d} \Omega^2 = - \mathrm{d}t^2 + (\mathrm{d}r-\partial_t R \mathrm{d}t )^2 + r^2 \mathrm{d} \Omega^2 \,, \qquad r = R(t,x)\,.
 \end{eqnarray}
The metric is degenerate in $(t,x)$ coordinates in the case of shell crossings, which means $\partial_x R =0$. As a result, the coordinate transformation from $(t,x)$ to $(t,r)$ is singular there. However, as can be seen from the solution \eqref{eq:general_solution}, all geodesics can pass through these degenerate points.

\subsection{Formation of trapped surfaces for the marginally and non-marginally bound case}
\label{sec:TrappSurf}
\noindent
In order to investigate the formation of trapped surfaces for a generic spherically symmetric spacetime described by the metric \eqref{eq:metric}, we can define two future-directed null vector fields which are normal to the shell with constant radius $R(t,x)=const$ via
\bq\label{null_expansion}
\partial_{\xi^+}=\frac{1}{\sqrt 2}\left(\partial_t+\frac{\sqrt{1 + \mathcal{E}}}{\partial_x R}\partial_x\right),\;\partial_{\xi^-}=\frac{1}{\sqrt 2}\left(\partial_t-\frac{\sqrt{1 + \mathcal{E}}}{\partial_x R}\partial_x\right).
\eq
If the radius of the shell shrinks along the radial null geodesics, i.e. $\xi^+=const$  and $\xi^-=const$, then a trapped surface is forming \cite{Hayward:1994bu}. In practice it is convenient to introduce the expansion parameters $\theta_\pm$ which are defined as
\bq
\theta_\pm%
=\frac{\sqrt 2}{ R}\left(\partial_t R \pm \sqrt{1 + \mathcal{E}}\right)\,,
\eq
where the sign refers to the expansion with respect to $\xi^{+}$ or $\xi^{-}$.\\

\noindent
From the expansion it is easy to see that the horizon is located at $\dot{R} = \pm 1$. There is a critical mass $M_c$ which is given by
\begin{eqnarray}
    M_c = \frac{8 \alpha}{3 \sqrt{3} G}  \,.
\end{eqnarray}
When $M(x) > M_c$ there exist both an inner horizon $\tilde\beta(x) -t = \pm h_{+}(x) $ and an outer horizon $\tilde\beta(x) -t = \pm h_{-}(x) $ s.t. $\theta_{+} \theta_{-} = 0$, where a trapped region is formed between $h_{-}$ and $h_{+}$ and an anti-trapped region is formed between $-h_{-}$ and $-h_{+}$. 
The other regions are untrapped. In the case that $M(x) \leq M_c$, there is no trapped region. Since the mass $M(x)$ depends on the radial coordinate $x$, it is possible that the trapped and untrapped region only appear in some restricted interval of $x$ instead of in the whole range $x>0$. This is the case, for example, inside a homogeneous dust cloud for which $M(x):=x^3 \epsilon$ with $\epsilon$ being the unit dust energy, as shown in \cite{Giesel:2022rxi}.

\subsection{Polymerized vacuum solution}
\label{sec:PolyVacuum}
\noindent
As stated before, the polymerized vacuum solution can be obtained by choosing $\mathcal{F} = R_s := 2 G M$ in \eqref{eq:general_solution} as in such case we have $\lambda =\widetilde{C}^{\Delta} = 0$ which means a vanishing dust density according to \eqref{eq:cov_eq}. Further choosing $\tilde\beta(x) = x$ the radial function $R$ is then with the help of equation \eqref{eq:general_solution} given by
\begin{eqnarray}
\label{eq:SolPolyVac}
R(x,t) = R(z) := \left( R_s \left( \frac{9}{4} z^2 +  \alpha^2 \right) \right)^{\frac{1}{3}} , \qquad z :=x-t\,.
\end{eqnarray}
This solution reproduces directly the Schwarzschild solution in LTB coordinates if we set $\alpha =0$. Moreover, as one can see such a solution is a stationary solution since $R$ only depends on $z$. Thus we have a Killing vector field $\partial_k= \partial_t + \partial_x$. It is straightforward to verify that $\partial_k$ is timelike outside the black hole horizon $z_{+}$. This gives a Birkhoff-like theorem which is analyzed in detail for such models in one of our companion papers \cite{GenBirkhoff}. We note that such a polymerized vacuum solution does not imply a vanishing $R_{\mu \nu}$ or $\phi$ in the mimetic model \eqref{eq:cov_action}. In contrast, the non-rotational dust density vanishes with a vanishing $\lambda$ but the mimetic model still involves the non-trivial coupling of $\phi$ in $T^{\phi}_{\mu\nu}$ which encodes quantum gravity effects. This can be seen from the curvature invariants%
\begin{eqnarray}
    \mathcal{R} = -\frac{96 \alpha^2}{\left(4 \alpha^2+9 z^2\right)^2} \, , \qquad \mathcal{R}_{\mu\nu\rho\sigma}\mathcal{R}^{\mu\nu\rho\sigma} = \frac{576 \left(160 \alpha^4-96 \alpha^2 z^2+27 z^4\right)}{\left(4 \alpha^2+9 z^2\right)^4}
\end{eqnarray}
which differ from the usual vacuum solution in the classical theory for which the scalar curvature  vanishes, that is $\mathcal{R} = 0$ since $\alpha=0$. The boundness of both curvature invariants at the bounce $z=0$ may suggest that the degeneracy introduced by $\partial_x R = 0$ at $z=0$ is a coordinate artifact for such polymerized vacuum solution. Moreover, 
the Einstein tensor $G_{\mu \nu}$ reads
\begin{eqnarray}
   G_{\mu\nu} \mathrm{d}x^{\mu} \mathrm{d}x^{\nu} =  - \frac{\mathcal{R}}{2} (-\mathrm{d}t^2 + (\partial_x R(z))^2 \mathrm{d}x^2)+ \mathcal{R} R^2 \mathrm{d}\Omega^2 =  - \frac{\mathcal{R}}{2} h_{ij}\mathrm{d}x^{i} \mathrm{d}x^{j}   + \mathcal{R} R^2 \mathrm{d}\Omega^2 \,,
\end{eqnarray}
with $h_{ij}$ the 2D metric on $\mathcal{M}_2$,
and hence the energy conditions, like the weak, the strong, the dominant and the null energy conditions still hold. As an example, for the strong energy condition, and a timelike vector $t^{\mu} = t_{h}{}^{\mu} + t_{\Omega}{}^{\mu}$ where we separate its angular part $t^{\Omega}$ along $(\theta,\phi)$ direction and the parts $ t_{h}{}^{\mu}$ in $\mathcal{M}_2$. We then have 
\begin{align}
    \mathcal{R}_{\mu \nu} t^{\mu} t^{\nu}  &= G_{\mu \nu} t^{\mu} t^{\nu} + \frac{1}{2} \mathcal{R} g_{\mu\nu}t^{\mu} t^{\nu} = - \frac{\mathcal{R}}{2}  t_{h}{}^{\mu} t_{h}{}_{\mu} + \mathcal{R} t_{\Omega}{}^{\mu} t_{\Omega}{}_{\mu}  + \frac{1}{2} \mathcal{R} \left(t_{h}{}^{\mu} t_{h}{}_{\mu} + t_{\Omega}{}^{\mu} t_{\Omega}{}_{\mu}  \right)\nonumber\\
    &= \frac{3}{2} \mathcal{R}t_{\Omega}{}^{\mu} t_{\Omega}{}_{\mu} > 0 \, . 
\end{align}
where we use the fact that $ t_{\Omega}{}^{\mu} t_{\Omega}{}_{\mu} > 0$ for the angular part. The other energy conditions can be seen by a similar method.\\

Since $R(z)$ is a smooth function across the bounce $z=0$, and by LTB condition we have $E^{\phi}(t,x) = R(z) \partial_z R(z)$, it is clear that $E^{\phi}$ thus $\det e$ has different signs before and after the bounce as $\partial_z R(z)$ does from the initial value problem imposed by \eqref{reduced equations of motion}. The degeneracy of the metric in LTB coordinates at the bounce $z=0$ is consistent with the fact that $\det e$ as a smooth function has different signs for $z>0$ and $z<0$. Clearly the solution gives smooth trajectories in the spherical symmetric phase space defined by \eqref{eq:cano_spherical}.

\subsection{Polymerized junction condition for the effective model}
\label{sec:PolyJuncCond}
\noindent
In this section we present junction condition for the effective model under consideration. It turns out that compared to the junction condition in the classical theory, this junction condition includes a fingerprint from the polymerization. For a given distributional energy momentum tensor of the form
\begin{eqnarray}
    T_{\mu \nu} = T_{\mu \nu}^{+} \Theta(n) +  T_{\mu \nu}^{-} (1 - \Theta(n)) + \sigma_{\mu \nu} \delta(n)
\end{eqnarray}
across a certain hypersurface indicated by $n =0$, the classical Israel junction condition reads
\begin{eqnarray}
    [K_{\mu\nu}] - \frac{1}{2} g_{\mu\nu} [K] = \frac{\kappa}{2} 
 \sigma_{\mu \nu},
\end{eqnarray}
where we introduce the notation $[f]:=f_{+}-f_{-}$ and $K_{\mu\nu}$ are corresponding second fundamental form and $\sigma_\mu\nu$ denotes the surface stress energy tensor.
With a continuum metric $[\gamma_{\mu \nu}] = 0$ across the junction surface, the junction condition can be derived from the  Einstein's equations in a distributional sense
\begin{eqnarray}
    G_{\mu \nu} = \frac{\kappa}{2} T_{\mu \nu} , 
\end{eqnarray}
or more specifically, 
\begin{eqnarray}
   \lim_{\epsilon \to 0} \int_{-\epsilon}^{\epsilon} G_{\mu \nu} dn = \frac{\kappa}{2} \lim_{\epsilon \to 0} \int_{-\epsilon}^{\epsilon} T_{\mu \nu} dn = \frac{\kappa}{2} \sigma_{\mu \nu} .
\end{eqnarray}
Given the fact that for an effective model and its underlying mimetic model we have modified Einstein's equations \eqref{eq:cov_eq}, we will obtain a modified junction condition instead of the classical one. One important remark is that, since in mimetic gravity we have the non-rotational dust-like energy-momentum tensor $2 \lambda \partial_{\mu} \phi \partial_{\nu} \phi$, the discontinuity of $\lambda$ is allowed which we denote as
\begin{eqnarray}
    \lambda = \lambda^{+} \Theta(n) + \lambda^{-} \left(1- \Theta(n) \right) + \delta \lambda \delta(n) \, .
\end{eqnarray}
A non-trivial $\delta \lambda$ on a timelike junction surface implies a shock solution.
~\\

\noindent
Note that due to the mimetic condition, we can relate all possible higher order derivatives of the scalar field $\phi$ with respect to $n$ to its derivative along the orthogonal direction in the junction surface. %
As a result, the distributional contributions stemming from derivatives of $\phi$ terms are all encoded in the derivatives of the metric components. If we further require a continuous induced metric across the junction surface, similarly to what one does for the classical junction condition, the distributional contributions in the modified Einstein tensor will be given explicitly in the derivatives of the possibly discontinuous extrinsic curvature terms. A detailed analysis is given in Appendix \ref{app:junction}, where we show that we obtain, compared to the classical case, a modified junction condition in the spherically symmetric spacetime. For a timelike junction surface, we have
\begin{align}\label{eq:junction_condition_general}
[ K_{\theta}{}^{\theta} ] &= -\frac{\kappa}{4} (-1)^m \sqrt{1-4 \alpha^2 Y^2} \sigma_{l}{}^{l}\\
\delta \lambda &= \frac{\kappa}{8 } \sigma_{l}{}^{l} \left(1-(-1)^m \sqrt{1-4 \alpha^2 Y^2} \right)\\
[K_{l}{}^{l}] &= - \frac{\kappa}{4  (-1)^m \sqrt{1-4 \alpha^2 Y^2} } \Big[\left(-4 \alpha^2 Y ((\sigma_{\theta}^{\theta}-3 \sigma_{l}^{l}) Y+\sigma_{l}^{l} X)-\sigma_{l}^{l}+2\sigma_{\theta}^{\theta}\right) \\
&\quad-\sigma_{l}^{l} (\partial_l \phi)^2 \Big(4 \alpha^2 Y \left(-\xi \left(X - 3Y \right)+(-1)^m Y \sqrt{1-4 \alpha^2 Y^2}\right) -(-1)^m \sqrt{1-4 \alpha^2 Y^2}+1\Big)\Big] \,  . \nonumber
\end{align}
where $(l,\theta,\phi)$ are coordinates in the induced space with the induced metric $d\hat{s} := h_{ll}^2 dl^2 + h_{\theta \theta}^2 d \Omega^2$ for spherical symmetric space time. The extrinsic curvatures $K_{ij}$ here are defined by
\begin{eqnarray}
    K_{ij} = \text{sign}(\det e) \frac{\partial_{{n}} h_{ij}}{2} \,.
\end{eqnarray}
And here we use the coupling of external matter sources $T_{\mu\nu}$ and the Mimetic energy density $\lambda$ with parity even coupling $\sqrt{-g}$ instead of $\det e$, otherwise a $\text{sign}(\det e)$ will appear in front of $\sigma_{\mu\nu}$ and $\delta \lambda$. 
As we can read off from the explicit form of the junction condition in \eqref{eq:junction_condition_general},  the polymerization effects are encoded in $X,Y$ and $\alpha$.  The classical junction condition is recovered by taking $\alpha =0$ and $m=0$ where the above junction condition becomes
\begin{align}
    [ K_{\theta}{}^{\theta} ] = - \frac{\kappa}{4}\sigma_{l}{}^{l}, \quad
\delta \lambda = 0, \quad 
[K_{l}{}^{l}] =  \frac{\kappa}{4} \left(\sigma_{l}{}^{l} -2 \sigma_{\theta}{}^{\theta} \right) \, .
\end{align}
There exists a special situation when $h_{ll}^2 =  (\partial_l \phi)^2$, or equivalently $\partial_n \phi =0$, in which case the condition on the extrinsic curvature becomes
\begin{align}\label{eq:junction_condition_dp0}
    [ K_{\theta}{}^{\theta} ]  &= - \frac{\kappa}{4} \left( \sigma_{l}^{l}-2 \delta \lambda \right)\\
   [K_{l}{}^{l}] &= \frac{\kappa}{4} (-1)^{m} \sqrt{1-4 \alpha^2 Y^2} (-2 \delta \lambda+\sigma_{l}^{l}-2\sigma_{\theta}^{\theta})
\end{align}
and $\delta \lambda$ can be chosen completely freely. If there is no external surface stress energy tensor present, that is $\sigma_{\mu \nu} = 0$, we have generally $[K_{\theta\theta}] = [K_{ll}] = \delta \lambda =0$, unless $\partial_n \phi =0$ in which case  non-vanishing $[K_{\theta\theta}]$ and $[K_{ll}]$ are allowed and determined by $\delta \lambda$. This is the case where only the dust surface density contributes. If $\partial_n \phi \neq 0$ and $\delta \lambda \neq 0$ there will be a discontinuity of the energy momentum tensor along the orthogonal direction.
~\\

\noindent
We remark that there also exists scenarios when the the multi-valued label $m$ across the junction surface is not the same. This is the case, for example, when we glue the post-bounce dynamics on the interior to some pre-bounce dynamics. In such situations, even if we have a continuous extrinsic curvature across the junction surface, the discontinuity in $m$ makes a difference. We will use the polymerized junction condition presented above to study the Oppenheimer-Snyder collapse model with a homogeneous dust cloud in detail in the next subsection.
\subsection{Effective Oppenheimer-Snyder dust collapse}
\label{sec:EffOppSny}
\noindent
As the LTB model presented in this work completely decouples in LTB coordinates, the junction condition for any dust shell trivializes in such coordinates and the dust collapse can be described by infinitely many decoupled cosmological dynamics with the dust energy density given at each  point $x$. This can be seen exactly from the general solution in \eqref{eq:general_solution} for the marginally bound case which admits any dust density profile specified by the choice of $\mathcal{F}(x)$. The usual Oppenheimer-Snyder (OS) dust collapse with a homogeneous dust profile inside the dust shell and vacuum outside can be rediscovered in such general solution with a non-smooth function $\mathcal{F}(x)=2G M(x)$ and $\tilde\beta(x)$ function given by
 \begin{eqnarray}\label{eq:os_collapse_f}
     \mathcal{F}(x) = 2 G \left[ x^3 E_0 \Theta(x_s - x) + x_s^3 E_0 \Theta(x-x_s)\right] , \qquad \tilde\beta(x) = \left( x-x_s \right) \Theta(x-x_s)
 \end{eqnarray}
The solution then takes the form
\begin{eqnarray}\label{eq:os_collapse_sol}
    R(t,x) &=& \Theta(x_s - x) x \left( 2G E_0 \left( \frac{9}{4}  (-t)^2+ 
\alpha^2 \right)\right)^{\frac{1}{3}}  + \Theta(x -x_s) \left( 2G M \left( \frac{9}{4}  (x-t -x_s)^2 + 
\alpha \right)\right)^{\frac{1}{3}}\nonumber \\
\end{eqnarray}
with $M= E_0 x^3$. We have that $R$ is continuous along $x$ at $x_0$ but not differentiable there. This leads to a discontinuous triad $E^{\phi} = R \partial_x R$ due to the LTB condition, but a continuous extrinsic curvature $[K]=0$ on the surface $x=x_0$, similar to what happens in the classical OS collapse that fulfills the junction condition.

\section{Comparison with existing models}
\label{sec:Comparison}
\noindent 
In this section we compare our results from subsection \ref{sec:MBEffMod} to subsection \ref{sec:EffOppSny} for the effective model derived in section \ref{sec:EffModfromCosm} with the existing models from  \cite{Kelly:2020uwj,Husain:2021ojz,Husain:2022gwp,Lewandowski:2022zce} and \cite{Fazzini:2023scu}.
~\\
\subsection{Areal gauge and effective dynamics}
\label{sec:ArialGauge}
\noindent
In order to compare the effective model from section \ref{sec:EffModel} with the model in \cite{Husain:2021ojz,Husain:2022gwp}, we want to explore the description of LTB spacetimes in a different set of coordinates. 
Classically generalized Gullstrand–Painlevé (GGP) coordinates have been shown to be a good choice to study the Hamiltonian dynamics of dust collapse \cite{Lasky:2006hq}. In order to transform the spherically symmetric metric written in terms of densitized triads in \eqref{eq:metric} to GGP coordinates, we have to employ the so called areal gauge $E^x =v^{\frac{2}{3}}  = x^2 :=r^2$.  Implementing the areal gauge, we can bring \eqref{eq:metric} into the GGP form
\be
\rmd s^2=-\rmd t^2+\L(t,r)^2\left( \rmd r^2 + N^r(t,r) \rmd t^2 \right)+r^2\lt[\rmd\theta^2+\sin^2(\theta)\rmd\varphi^2\rt] .
\ee 
Now we want to repeat this procedure in the effective model given by equation \eqref{eq:effectiveHamiltonian} to get to the GGP type of solutions of that model. Note that gauge fixing the diffeomorphism constraint $C_x$ with respect to the areal gauge condition, we can completely eliminate the radial component of the extrinsic curvature $K_x$ and the triad $E^x$. It turns out that the corresponding Dirac bracket is the same as the Poisson bracket for the remaining set of variables which is $\poissonbracket{K_\phi}{E^\phi}_D = \poissonbracket{K_\phi}{E^\phi}$. Nevertheless, we will study here diffeomorphism invariant solutions with $E^x  = r^2$ using the full set of equations of motion. As we require the areal gauge fixing to be stable under the effective dynamics generated by the primary Hamiltonian in \eqref{eq:effectiveHamiltonian}, we have that the Hamilton's  equation for $E_x$ has to vanish on-shell. We can write the dynamical equation $\partial_t E^x$ in the areal gauge as
\be 
0 = \partial_t r^2 = r \left(-2 N^r+\frac{r \sin \left( \frac{2 \alpha K_\phi}{r}\right)}{\a}\right)
\ee 
which implies that the shift vector has to be chosen for this kind of solution as
\be\label{solNx_x^2}
N^r = - \frac{r \sin \left( \frac{2 \alpha K_\phi}{r}\right)}{2 \a}\,.
\ee 
We realize that the factor of $2$ in the argument of the sine function in the shift vector consistently follows from the stability with respect to the effective dynamics encoded in \eqref{eq:effectiveHamiltonian} and is thus a consistent gauge fixing of the effective model along the lines of \cite{Giesel:2021rky}.
~\\
In this gauge the physical Hamiltonian assumes up to boundary terms the form
\begin{equation}
   H^\Delta_{phys}= -\frac{1}{2 G} \left[ \frac{E^\phi}{x \alpha^2}\partial_x\qty(x^3 \sin^2\qty(\frac{\alpha K_\phi}{x} )) + \frac{E^\phi}{x}  + \frac{x}{E^\phi}  \right]\,.
\end{equation}
The equations of motion generated by this Hamiltonian can be computed using the relation $\poissonbracket{K_\phi}{E^\phi}_D = \poissonbracket{K_\phi}{E^\phi} = G \delta(x,y)$ and the result is
\be 
\partial_t E^{\phi} &=&  -r^2 \partial_r \left( \frac{E^{\phi}}{r} \right)  \frac{\sin \left(\frac{2 \alpha K_\phi}{r}\right)}{2 \alpha} \label{ed_eom_eb}\\
\partial_t K_\phi
&=&- \frac{1}{2 r \alpha^2}\partial_r \left( r^3 \sin ^2\left(\frac{\alpha K_\phi}{r} \right) \right)-\frac{1}{2 r}  +\frac{r}{2{E^{\phi}}^2} \label{ed_eom_b}\,. \nonumber
\ee 
Note that the equation of motion for $K_x$ in the unreduced system does not contain any independent information but reduces to a consistent equation with the vanishing of diffeomorphism constraint. Comparing these results to the work in \cite{Husain:2021ojz,Husain:2022gwp}, we see that by identifying $E^{\phi}$ with $E^b$ and $\gamma K_{\phi}$ with $b$, these equations of motions do exactly coincide.\\

\noindent
This is a strong evidence that the effective model found in \cite{Husain:2021ojz,Husain:2022gwp} is actually equivalent to the model presented here and can be derived from the underlying effective spherically symmetric model that can be understood as a, with respect to the areal gauge, gauge unfixed version of the model in \cite{Husain:2021ojz,Husain:2022gwp}. However, it will be shown in subsection \ref{sec:CompPolyVac} that the areal gauge is not the optimal choice to obtain a good coordinate system in which the model can be analyzed. For instance  working in LTB coordinates, a bouncing vacuum solution can be found, however this is not the case when implementing the areal gauge. As a consequence, this puts the presence of shock waves as found in \cite{Husain:2021ojz,Husain:2022gwp} into a new perspective because starting from a dust collapse described in LTB coordinates, the effective model involves a symmetric bounce instead of a shock wave.\\

\noindent
It is easy to obtain the solution in GGP with areal gauge by a coordinate transformation  $r=R(x,t)$ from LTB coordinate, which is given by
\begin{eqnarray}
    \rmd s^2 = - \rmd t^2 +( \rmd r - (\partial_t R(t,x)) \rmd t)^2 + r \rmd \Omega^2 \, ,
\end{eqnarray}
where we use $\rmd r = \partial_x R(t,x) \rmd x + \partial_t R(t,x) \rmd t$.
Thus we have
\begin{eqnarray}
    N^r = - \partial_t R(t,x) = - \text{sign}( \partial_t R(t,x)) \sqrt{\frac{2 G M(x)}{r}\left( 1 - \frac{\alpha^2}{r^2} \frac{2 G M(x)}{r} \right) } \, ,
\end{eqnarray}
where we use \eqref{Friedmann from homogeneous reduction}.
Here from such coordinate transformation we have naturally a sign assigned to $N^r$, s.t., before the bounce where $\partial_t R > 0 $ we have $N^r < 0 $ and after the bounce where $\partial_t R < 0 $ we have $N^r > 0 $. $N^r$ is actually a smooth function across the bounce, in contrast from the models found in \cite{Husain:2021ojz,Husain:2022gwp} where only one sign is allowed for the whole evolution in GGP with areal gauge. We will see later on that such sign plays an important role for the junction condition in gravitational collapse in section \ref{sec:CompOSJuncCond}. We finally remark that, starting directly in GGP coordinate with areal gauge, one has
\begin{eqnarray}
    2 \alpha b = \sin^{-1}\left(-\frac{2 \alpha N^r}{r} \right) ,
\end{eqnarray}
As a result, while we have a continuous $b$ solution across the bounce in LTB coordinate, in \cite{Husain:2021ojz,Husain:2022gwp,Munch:2021oqn} only one branch of $b$, e.g. $\alpha b < \pi/2$ or $\alpha b > \pi/2$ is allowed. This is the reason why in \cite{Husain:2022gwp}, for the polymerized vacuum solution $\alpha b < \pi/2$ is found, which removes the possibility of the presence of a bounce in forward time direction for this solution.

We note that in the model presented in \cite{Husain:2021ojz,Husain:2022gwp} the absolute value of $|E^{\phi}|$ is presented in the Hamiltonian, namely they use the parity even coupling $\sqrt{-h}$ instead of the parity odd $\det e$. However, in the $r$ coordinate where only one branch of $b$ is allowed, $E^{\phi}$ will not change its sign, thus such difference can not be observed. 

\subsection{Comparison of polymerized vacuum solutions}
\label{sec:CompPolyVac}
\noindent
First we make use of the fact that we know the underlying covariant mimetic model of the polymerized vacuum model presented in section \ref{sec:PolyVacuum} and construct a suitable coordinate transformation which shows that we can rediscover the form of the metric obtained in the models in \cite{Kelly:2020uwj,Lewandowski:2022zce}. For this purpose we define the following coordinate transformation
\begin{eqnarray}
    && (t,x) \to (\tau,z), \quad \tau = t + \tau(z), \qquad \tau(z):=\int_{z_0}^z \mathrm{d}z' \frac{ (R')^2}{(R')^2 - 1}, \nonumber
\end{eqnarray}
with $R(z)$ the solution given in \eqref{eq:SolPolyVac} that we denote as generalized Schwarzschild-like coordinates because they merge into Schwarzschild coordinates if we consider the classical case.  This enables us to transform the metric in \eqref{eq:metric} with $N^x=0$ to the following form
\begin{eqnarray}
\label{metrictz1}
 \mathrm{d} s^2 = - ( 1-(R'(z))^2) \mathrm{d} \tau^2 +\frac{R'(z)^2}{ 1-(R'(z))^2} \mathrm{d}z^2 + R(z)^2 \mathrm{d} \Omega^2\, .
\end{eqnarray}
By further defining $r = R(z)$ 
the metric can be rewritten as
\begin{eqnarray}
    \mathrm{d}s^2%
    &=&- A(r) \mathrm{d} \tau^2 + \frac{1 }{A(r)} \mathrm{d} r^2 + r^2 \mathrm{d} \Omega^2
\end{eqnarray}
with $A(r)$ given by 
\begin{equation}\label{eq:metric_in_r}
   A(r) = 1 -  \frac{2 G m_s}{r} \left( 1 -\frac{\alpha^2    }{r^2 }\frac{2  G m_s}{ r} \right). 
\end{equation}
This metric has exactly the same form as the one obtained in \cite{Kelly:2020uwj,Lewandowski:2022zce}. Thus this model provides a consistent $(1+1)$-dimensional model beyond the junction conditions used in \cite{Lewandowski:2022zce}. However, we realize that the metric in \eqref{eq:metric_in_r} holds for the whole $r>0$ region except the horizons where coordinate singularities are present, but the coordinate transformation is only well defined for $r=R(z) \geq R|_{\min}$ where $R(z)$ is monotonic in $z$. The extension of \eqref{metrictz1} beyond this region is not a solution defined in the covariant mimetic model of \eqref{eq:cov_action}. 
Actually a detailed analysis of the equations of motion of the covariant action \eqref{eq:cov_action} shows that the corresponding vacuum solutions, that we denoted as polymerized vacuum solutions, are given by $A(r)$ for $r \geq r_{\min}$ only (see Appendix \ref{app:cov_sol} for more details on the derivation of the solutions).\\

\begin{figure}[h]
    \centering 
    \begin{subfigure}[b]{0.46\textwidth}
    \includegraphics[width=0.9\linewidth]{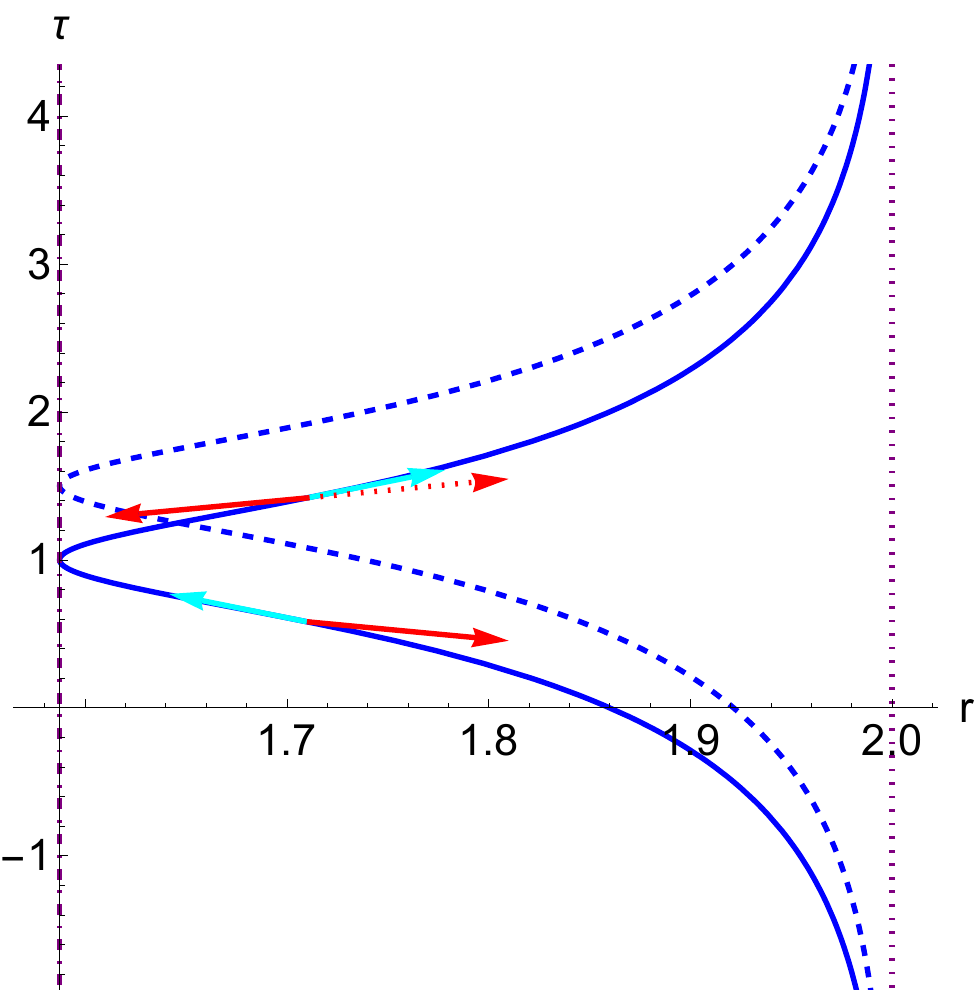}
    \caption{$x=1$ and $x=3/2$ curves plot in $(\tau,r)$ coordinates. They will intersect each other. Blue and orange line coincide in this case.}
    \end{subfigure}
    \begin{subfigure}[b]{0.46\textwidth}
    \includegraphics[width=0.9\linewidth]{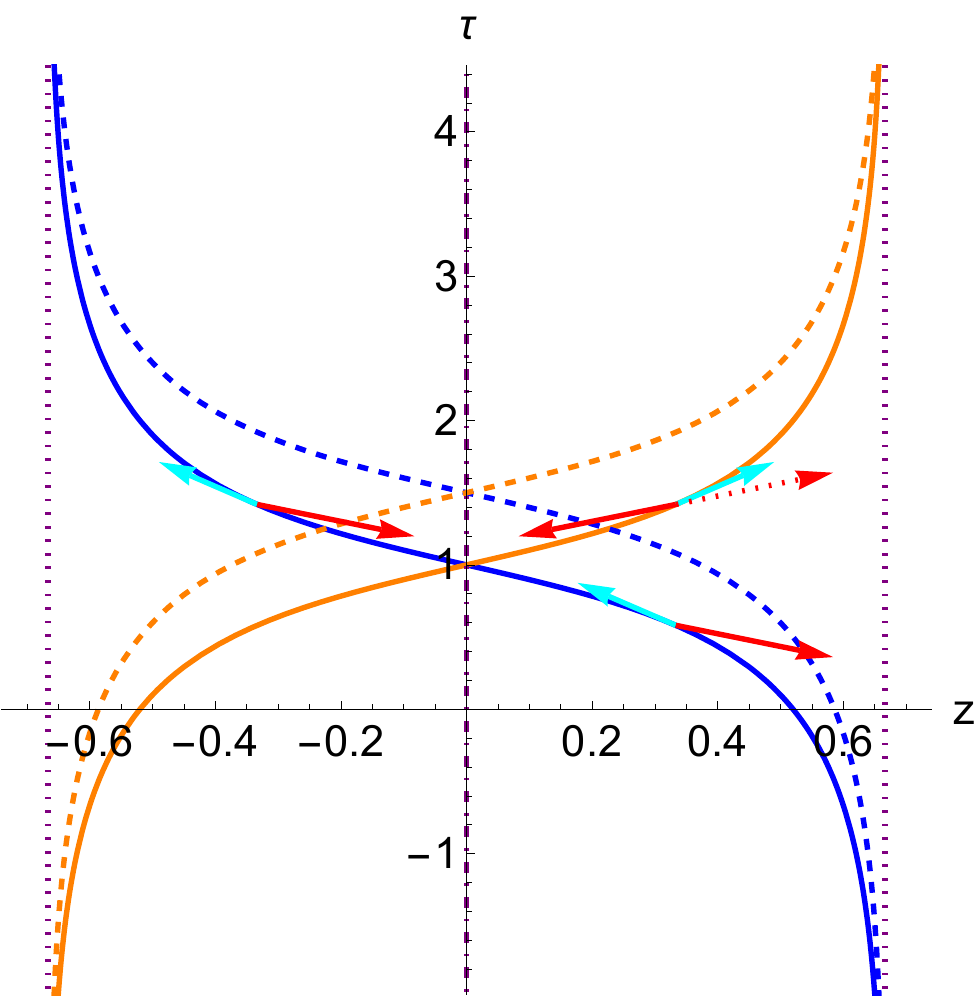}
    \caption{$x=1$ and $x=3/2$ curves plot in $(\tau,z)$ coordinates. Blue lines will not intersect with blue lines but will intersect with orange lines.  }
    \end{subfigure}
    \begin{subfigure}[b]{0.56\textwidth}
    \includegraphics[width=0.9\linewidth]{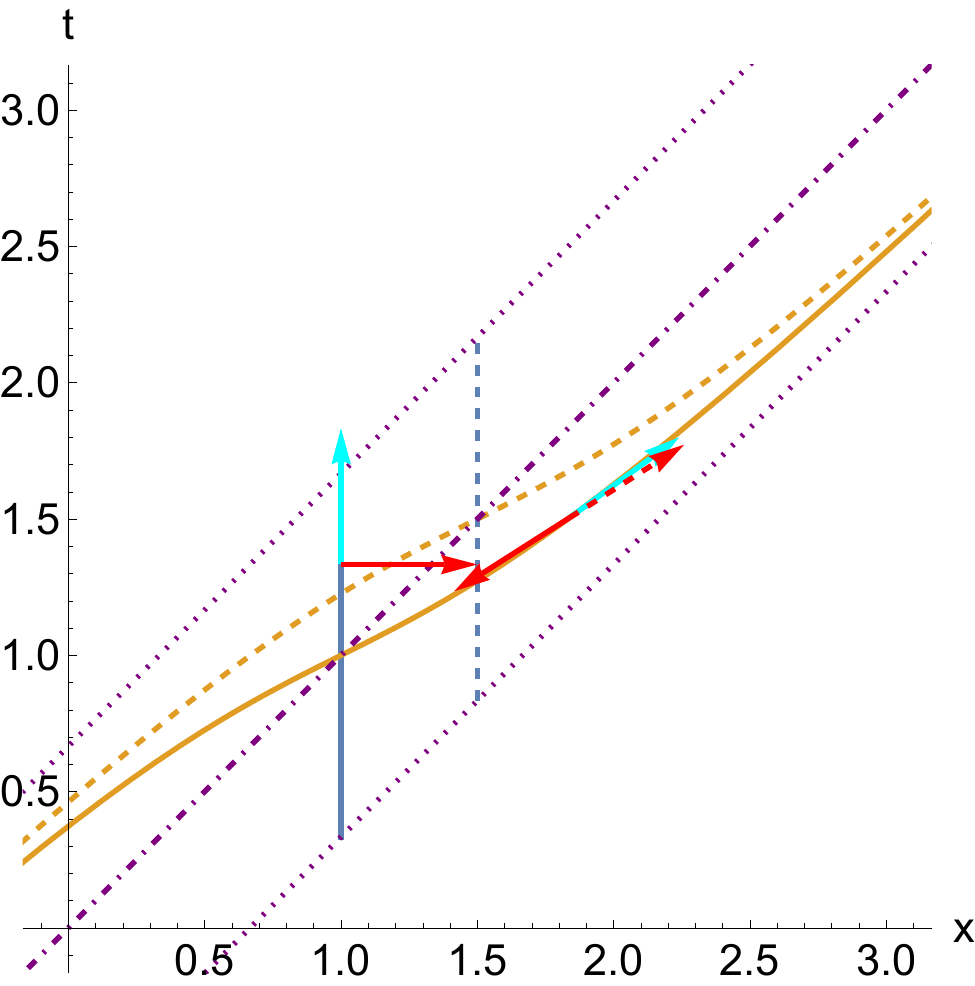}
    \caption{$x=1$ and $x=3/2$ curves plot in $(\tau,z)$ coordinates. Blue lines will not intersect with blue lines but will intersect with orange lines.  }
    \end{subfigure}
    \caption{Blue line: $x=1$ (thick) and $x=3/2$ (dashed) curves in different coordinates. Orange line: The corresponding curve where we identify $(\tau,z) \to (\tau,-z)$. The dashed purple lines mark the position of the inner horizon where the dot dashed purple lines mark $z=0$ where the bounce happens. The cyan vectors mark the tangent direction of the corresponding worldlines and the red vectors are the normal vectors pointing increasing $x$ directions. The dashed red vectors are the normal vectors mark the direction where a consistent orientation with blue lines can be given in the $r>r_{min}$ or $z>0$ region. %
    }
    \label{fig:mapping-coord}
\end{figure}

\noindent
Due to the fact that $R(z)$ is actually a function of $z^2$ from \eqref{eq:SolPolyVac}, the $(\tau,z)$ and $(t,x)$ coordinates are in a $2$-to-$1$ correspondence with the coordinates $(\tau,r)$ for solutions of the equations of motion of the covariant action in \eqref{eq:cov_action}. Starting from $(\tau,r)$ coordinates this then introduces an identification map in $(\tau,r)$ between points $(\tau,z)$ and $(\tau,-z)$, by reversing $z = \pm R^{-1}(r)$ where $R^{-1}$ is the inverse function of $R(z)$ which has a branch cut due to the square root.
From the coordinate transformation between $(\tau,z)$ and $(\tau,x)$, this corresponds to $(t,x)$ and $(t + 2\tau_0(x-t), -x + 2t+ 2\tau_0(x-t) )$ as well, where the details are shown in Appendix \ref{app:coord_trans}. The corresponding identification is illustrated in figure \ref{fig:mapping-coord}. 
We want to point out that while the identification map introduces a singular derivative at $z=0$, the solution is still well defined on it as on the $z=0$ surface the metric is also degenerate.
~\\

\noindent
However, the flipped patch from $(\tau,z)$ to $(\tau,-z)$ has a different orientation towards increasing $t,x$ direction, as can be seen from the tangents and normals along the geodesics $x=const$ which are world lines of the mimetic field $\phi$. This implies the orientation in terms of $(t,x)$ is changed if we stay with $(\tau,z)$ before the bounce but switch to $(\tau, -z)$ after the bounce. Such observation is consistent with the fact that in $(t,x)$ coordinates after the bounce we have a sign change of $E^{\phi}$ and thus of $\det(e)$. Such a scenario is also observed directly in $(\tau,r)$ coordinates before and after the bounce. As a result, staying in $(\tau,z)$ or $(t,x)$ coordinates (blue lines in figure \ref{fig:mapping-coord}) as implied by our solution from the initial value problem introduce a different global structure of the spacetime than the one in $(\tau,r)$ coordinates. In $(\tau,r)$ coordinates after the bounce the interior region of the geodesic $x=const$ for increasing $r$ is not the direction of increasing $x$. Moreover, in $(t,x)$ coordinates, all $x=const$ geodesics will not intersect with each other, but this is not the case in $(\tau,r)$ coordinate. In $(\tau,r)$ coordinate, e.g. a $x=x_0 = const$ geodesic will intersect with all $x> x_0 =const$ after the bounce. As $x=const$ are the worldline of our clock field which is the mimetic field, this introduces the discontinuity of our clock field thus is not favored by our model. %
We realize, in the $(\tau,r)$ coordinates, due to this correspondence, the solution is both a solution of the mimetic model where we couple with either $\sqrt{-h}$, which is an even function under a parity transformation, or choose for the coupling $\det(e)$ which is odd function under parity transformation. Consequently, the solution is insensitive to the time direction $z$ in LTB coordinates after the bounce, which is consistent with the time-reversed solution in \cite{Munch:2021oqn} for models in GGP with areal gauge. In the model where we choose a coupling that is even under parity transformation, the solution will only cover the region $z \geq 0$ (or $z \leq 0$). Therefore, in a collapsing scenario there will be two different junction conditions available, which we will discuss in the following subsections.
~\\

\noindent
The fact that the solution in $(\tau,r)$ coordinates is in a 2-to-1 map may also change the value of $b$ in the corresponding solutions. In contrast to the discontinuous function $b(t,x)$ presented in \cite{Husain:2022gwp} this function is continuous here. The continuity here is clear from \eqref{reduced equations of motion} where $b$ is a monotonic function. The reason that $b$ is monotonic is closely related to the fact that we have a  Hamiltonian that is odd under parity transformations. 
Here the continuous $b$ function across the bounce is given by
\begin{eqnarray}
    2 \alpha b(t,x) = \sin^{-1}_{(m)}(2 \alpha Y)
\end{eqnarray}
with $2 \alpha b >\pi$ after the bounce. 
However, in the $(\tau,r)$ coordinates used in \cite{Husain:2022gwp}, we must identify $2 \alpha  b > \pi$ with $2 \pi - 2 \alpha b < \pi$, as for an observer along the $x>x_0$ geodesic before the bounce will meet the observer at $x=x_0$ after the bounce, thus their $b$'s need to be identified in order not to have a multi-valued $b$. However, from the relation between $b$ and $Y$, we see immediately that for the two cases, the sign of $Y$, which is related to the sign of $\partial_t E^x = -\partial_x E^x$ thus also the sign of $E^{\phi}$, is switched. This is a strong evidence that actually if we switch to the orange line in figure \ref{fig:mapping-coord} after the bounce, we are actually moving backward in time. 
~\\

\noindent
We finally remark that, in order to make the solution in $(\tau,r)$ coordinates to have the same global structure as our solution presented in LTB coordinates, one must break such 2-to-1 correspondence and thus have two patches of $(\tau,r)$ coordinates before and after the bounce. This is exactly the case considered in \cite{Munch:2020czs,Munch:2021oqn} where in that case a junction condition is necessary due to the discontinuity of $N^r$ or $b$ in GGP coordinates with areal gauge. In the model presented here we have already seen in the previous section that such a junction condition is not necessary and we have a smooth metric across the bounce by solving the initial value problem from \eqref{reduced equations of motion}. The global extension will be similar to the one given in \cite{Munch:2021oqn} with an infinite extended causal diagram.
\subsection{Oppenheimer-Snyder collapse and junction conditions}\label{compare_junction}
\label{sec:CompOSJuncCond}
\noindent 
In this subsection we will compare our result for the Oppenheimer-Snyder collapse with the results obtained in  \cite{Husain:2022gwp,Lewandowski:2022zce} and \cite{Fazzini:2023scu}. One important difference is that, our solution \eqref{eq:os_collapse_sol} is the solution of \eqref{reduced equations of motion} with the initial value given at certain $t=const$ surface. However, the solution of \cite{Husain:2022gwp,Lewandowski:2022zce} and \cite{Fazzini:2023scu} are not obtained directly from an initial value problem but junction condition are used. Thus to compare we will also formulate our solution using the polymerized junction condition presented in section \ref{sec:PolyJuncCond} and study whether the solutions presented in \cite{Husain:2022gwp,Lewandowski:2022zce,Fazzini:2023scu} are allowed in our model.
~\\
~\\
From section \ref{sec:PolyJuncCond}, we realize that when there is no surface energy density present, the classical junction condition, that any geodesic whose extrinsic curvature is continuous across the surface it defines, can still be used to glue the spacetime. However, when there is a non-vanishing surface stress energy tensor, the junction condition gets modified in general and polymerization effects contribute. This is the case, e.g. in \cite{Husain:2022gwp} where a weak solution is allowed. This is the underlying reason why the discontinuity and the surface energy density appear and one obtains a shock solution.
~\\
~\\
In our solution we have clearly $\partial_n \phi =0$ on the  $x=x_s=const$ shell.
A simple calculation of solution \eqref{eq:os_collapse_sol} along the shell implies $K_{ll} = 0$ and $K_{\theta \theta} = \frac{\partial_x R}{ \partial_x R} = 1$ for both the interior and the exterior. Thus the junction condition \eqref{eq:junction_condition_dp0} is satisfied without any surface stress energy tensor. 
~\\
~\\
In \cite{Husain:2022gwp,Lewandowski:2022zce,Fazzini:2023scu} the exterior vacuum solution are all formulated in $(\tau,r)$ coordinates instead of LTB coordinates $(t,x)$ which we use here. 
As we analyzed in \ref{sec:CompPolyVac}, $(\tau,r)$ coordinates actually have a different global structure than the polymerized vacuum solution presented in \ref{sec:PolyVacuum}.  However, 
due to the 2-to-1 correspondence between $(t,x)$ and $(\tau,r)$ coordinates, there exists another possible junction condition for the gluing except the trivial one \eqref{eq:os_collapse_f} in the LTB coordinate. As illustrated in figure \ref{fig:mapping-coord}, this corresponds to when we first follow the blue line before the bounce and then the orange line after the bounce such that we obtain the same global structure as for $(\tau,r)$ coordinates in the $(t,x)$ coordinates. Since both lines are geodesics, the gluing conditions for the extrinsic curvatures are satisfied with $\sigma_{\mu\nu} = \delta \lambda = 0$. However, we remark that when one follows the orange line after the bounce, the mimetic field $\phi$ along the orange line does not satisfy $\partial_n \phi = 0$ along its normal direction. This can be easily seen from the $(t,x)$ coordinates in figure \ref{fig:mapping-coord}, as the orange line is no longer the world line of the mimetic field $\phi$. From the mimetic condition this implies a discontinuity on the left and right of the junction surface of the mimetic field along the tangent direction. The detailed analysis is given in Appendix \ref{app:junction_noshock}.
~\\

\noindent
Moreover, as analyzed in \ref{sec:CompPolyVac}, in $(\tau,r)$ coordinate we must allow the identification of $2 \alpha b > \pi$ with $2 \pi - 2 \alpha b < \pi$. This turns out to be the junction of the $2 \alpha b_{in} > \pi$ in the homogeneous dust collapse interior with the $2 \alpha b_{ex} < \pi$ for the vacuum exterior. In \cite{Husain:2022gwp}, to overcome such discontinuity between $b$, a shock solution is introduced. The shock solution presented in \cite{Husain:2022gwp} is a shock generated only from the dust-like density $\delta \lambda$ without introducing external sources. From the polymerized junction condition \eqref{eq:junction_condition_general} we notice that a non trivial $\delta \lambda$ without external $\sigma_{\mu\nu}$ is only possible in the case of $\partial_n \phi =0$ on the junction surface. This is the case for the homogeneous interior. However, for the exterior, this implies that the junction surface must be the world line of the mimetic field $\phi$ thus going back to our solution which has $\delta \lambda = 0$. Consequently, the shock solution in \cite{Husain:2022gwp}  violates the smoothness of the mimetic field in the covariant model. The detailed analysis is given in Appendix \ref{app:junction_shock}.

\section{Conclusions}
\label{sec:Concl}
\noindent
This work is build on one of the results of our companion paper \cite{LTBStabiliyPap1}, namely that if we consider effective LTB models for which the corresponding physical Hamiltonian densities are conserved, then those class of models consists of infinitely many decoupled effective LQC models along the radial axis. This allowed us to construct the underlying effective spherically symmetric model in section \ref{sec:EffModel} and link it to the associated covariant mimetic gravity model that we presented in section \ref{sec:CovLagrangian}.  
Our analysis led us to new insights not only concerning the underlying effective spherically symmetric and its corresponding mimetic model but we could also generalize existing results. 
~\\
~\\
First our strategy provides a systematic procedure on how effective spherically symmetric models can be constructed from LQC models under certain assumptions the effective spherically symmetric models needs to satisfy. Furthermore, by considering the associated covariant mimetic model it turns out that in order to match the decoupled set of LQC models we need to consider a coupling by means of the $\det(e)$ in the mimetic model instead of the usual coupling involving $\sqrt{-h}$. The reason for this is that $\det(e)$ is a pseudo-scalar under parity transformation. This carries over to the form of the effective physical Hamiltonian and as a consequence such models allow one of the triads to change the sign before and after the bounce. Next, we consider the effective marginally bound case and obtain an analytical solution extending those presented in \cite{Giesel:2022rxi} and \cite{Fazzini:2023scu} because here the model can also account for inhomogeneous dust profiles. 
~\\
~\\
Furthermore, the underlying covariant mimetic model and its corresponding modified Einstein's equations were taken as a guiding principle to derive polymerized junction conditions for the model, whereas to the knowledge of the authors such a derivation is not present in the literature yet. It turn out that when there is no surface stress energy tensor present, the classical junction condition can still be used, whereas for a non-vanishing stress energy tensor polymerization effects contribute.  As an application of the polymerized junction condition we investigated the effective Oppenheimer-Snyder dust collapse in section \ref{sec:EffOppSny} and we saw that the junction condition trivializes in LTB coordinates due to the fact that the effective LTB model consists of infinitely many decoupled LQC models along the radial coordinate. In contrast to the existing models the analytical solution for the homogeneous dust collapse presented in subsection \ref{sec:EffOppSny} is a solution of an initial value problem for these decoupled models and therefore in LTB coordinates junction conditions are not a mandatory part of the effective dust collapse model. 
~\\
~\\
Another aspect where the generalized analysis of effective dust collapse models of this work is useful is in the comparison with existing models that we discussed in section \ref{sec:Comparison}.  The main reason for this is that we have on the one hand the underlying covariant mimetic model available that allows us to consider coordinate transformations including also transformations of the temporal coordinate and on the other hand we have the underlying spherically symmetric model involving no areal gauge fixing yet that can be restricted to its LTB sector in effective dynamics. 
As there is an ongoing interest in the question whether such an effective dust collapse admits shock solutions and if so under which assumptions this was one of our foci in the comparison. Our results can be summarized as follows: first, given the underlying effective spherically symmetric model we showed in subsection \ref{sec:ArialGauge} that this model can be understood as the with respect to the areal gauge unfixed counter part of the model introduced in \cite{Husain:2021ojz,Husain:2022gwp}. As we presented the corresponding covariant mimetic model for the effective spherically symmetric model, with this result we can also link it to the model in \cite{Husain:2021ojz,Husain:2022gwp}. Next we considered the polymerized vacuum solution presented in subsection \ref{sec:PolyVacuum} and showed that by means of the application of a suitable coordinate transformation we can transform the effective metric into the form of the models in \cite{Kelly:2020uwj,Lewandowski:2022zce}. As the consequence the polymerized vacuum solution in this work provides a consistent $1+1$-dimensional model beyond the junction condition used in \cite{Lewandowski:2022zce}. From the perspective of the covariant model we showed that using generalized Schwarzschild-like coordinates $(\tau,r)$ the polymerized vacuum solution is only defined in the region where $r$ is larger than the minimal radius.
~\\
~\\
In subsection \ref{sec:CompPolyVac}, we compared the different coordinates used for the vacuum solution. We showed that there is a 2-to-1 correspondence between generalized Schwarzschild-like coordinates $(\tau,r)$ and the LTB coordinates, because the radius of any sphere bounces and is an even function in the LTB coordinates. Therefore, the $(\tau,r)$ coordinates only cover half of the spacetime. We also showed that a single $(\tau,r)$ coordinate patch cannot reproduce the same spacetime as our polymerized vacuum solution, because it has discontinuities of the mimetic field that are not present in our solution. Moreover, it cannot capture the signature changes of $\det e$ presented in the solution discussed in subsection \ref{sec:PolyVacuum}. To match this result, one needs two $(\tau,r)$ patches, one before and one after the bounce, similar to \cite{Munch:2020czs,Munch:2021oqn} where the junction condition is considered to glue the two patches. This demonstrates that the covariant model presented in subsection \ref{sec:CovLagrangian} provides a consistent model with a polymerized vacuum solution based on an initial value problem, beyond the junction condition used in \cite{Munch:2020czs,Munch:2021oqn}.

~\\
A further insight obtained from the covariant theory related to the presence of shock solutions is that our results in section \ref{sec:CovLagrangian} show that we need to work with a covariant model where the coupling is formulated in terms of $\det(e)$ instead of $\sqrt{-h}$ in order to reproduce the correct dynamics of the infinitely many decoupled LQC models. This has the consequence, that in the marginally bound case we have an effective dust collapse model where in LTB coordinates the elementary variable related to the extrinsic curvature is continuous allowing a symmetric bounce in contrast to the model in \cite{Husain:2021ojz,Husain:2022gwp} where a shock solution after the bounce is present.
As an application, in section \ref{sec:CompOSJuncCond} we compare our results on the effective Oppenheimer-Snyder collapse with those in \cite{Husain:2022gwp,Lewandowski:2022zce} and \cite{Fazzini:2023scu}. The latter are formulated in terms of junction conditions and do not arise directly from an initial value problem as in our case, so we have used the polymerized junction condition derived in section \ref{sec:PolyJuncCond} to formulate the model here and perform this comparison. For the polymerized model presented in this work, the mimetic field $\phi$ satisfies $\partial_n\phi=0$ on $x=x_s=$const shells, where $n$ denotes the normal direction, and with this specification the model satisfies the corresponding polymerized junction conditions without any surface energy terms. Considering this junction condition there are no shock solutions in LTB coordinates, but the model includes a symmetric bounce.
~\\
~\\
If we consider the Schwarzschild-like coordinates for the extrinsic curvature part as used in \cite{Husain:2022gwp,Lewandowski:2022zce,Fazzini:2023scu}, then, as our discussion in section \ref{sec:CompOSJuncCond} reveals, there exists a second possible junction condition for gluing due to the 2-to-1 correspondence between LTB and Schwarzschild-like coordinates. For this possibility, the extrinsic curvature conditions are still satisfied with vanishing surface energy density. However, in this case the condition $\partial_n\phi=0$ is violated after the bounce, leading to a different model than the one presented here, since one of the assumptions of the model is omitted. From the perspective of the polymerized junction condition we can understand the shock solution in \cite{Husain:2021ojz,Husain:2022gwp} to be generated only from a dust-like density with still vanishing external surface stress energy tensor. As our analysis has shown, this requires $\partial_n\phi=0$ on the junction surface. Thus, the shock solution presented in \cite{Husain:2022gwp} violates the smoothness of the mimetic field and thus is not viable from this perspective. \\

\noindent
In the recent work \cite{Fazzini:2023scu} it was discussed that the presence of shock solutions in the model in \cite{Husain:2021ojz,Husain:2022gwp} can be understood as a coordinate artifact using GGP coordinates and is in addition related to some discontinuity in the dust chosen as a temporal reference field in \cite{Husain:2021ojz,Husain:2022gwp}. From the insights of the underlying covariant model, we can understand the temporal gauge fixing used in \cite{Husain:2021ojz,Husain:2022gwp} in terms of the mimetic field, which, unlike the (non-rotating) dust, has a more complicated coupling with gravity, including higher order derivatives. We thus find that we have to abandon the assumption of the smoothness of the mimetic field in order to get into situations where shock solutions are admissible. Furthermore, since the mimetic field plays the role of the temporal reference field in the effective model, the difference between the two models can also be understood as different properties that the temporal reference field must fulfill. 
~\\

\noindent
As one of the next step in future work we plan to investigate effective LTB models with inhomogeneous dust profiles and the extension of the study to other effective models, e.g. the underline decoupled model is the effective LQC model with an asymmetric bounce \cite{Yang:2009fp,Liegener:2019ymd,Han:2018fmu}.

\section*{Acknowledgements}
This work is supported by the DFG-NSF grants PHY-1912274 and 425333893 and NSF grant PHY-2110207. H.L. acknowledges Beijing Normal University for the hospitality during his visits while this work was completed.

\bibliographystyle{jhep}
\bibliography{refs}

\onecolumngrid
\appendix
\vspace{3em}
\section{Coordinate transformations}\label{app:coord_trans}
\noindent We want to start this analysis with the general solution in the marginally bound case defined by $\mathcal{E}(x) = 0$. The LTB metric can be written in this case as
\begin{eqnarray}
    ds^2 = - dt^2 + (\partial_x R(x,t))^2 dx^2 + R(x,t)^2 d \Omega^2\,,
\end{eqnarray} 
where the radial function $R(x,t)$ is given by 
\begin{eqnarray}
    R(x,t)= \mathcal{F}(x)^{\frac{1}{3}} \tilde{R}(z) , \qquad \tilde{R}(z):=\tilde{R}(\tilde\beta(x) - t):=\left( \frac{9}{4} (\tilde\beta(x) - t )^2 +  \alpha^2 \right)^{\frac{1}{3}}\,.
\end{eqnarray}
Further we have $\mathcal{F}(x) = 2 G M(x) := 8 \pi G \widetilde{C}(x)$ and $\tilde\beta(x)$ is an arbitrary function.\\

First we want to introduce the coordinate transformation $(t,z) \to (\tau,z)$ via
\begin{eqnarray}
    \tau = t + \int_{z_0}^z dz' \frac{ (R')^2}{(R')^2 - 1}\,,
\end{eqnarray}
which is monotonic in $t,z$ in each region bounded by the horizon, where $R' = \pm 1$, and the asymptotic boundary at $z \to \pm \infty$. The coordinate $\tau(t,z)$ is smooth for both $t$ and $z$ and the Jacobian of the coordinate transformation is given by
\begin{eqnarray}
    \frac{\partial (\tau,z)}{\partial (t,x)} = \left(\begin{array}{cc}
        1 + \frac{(R')^2}{1-(R')^2}&  -\frac{(R')^2}{1-(R')^2} \\
        -1 & 1
    \end{array} \right)\,,
\end{eqnarray}
which is clearly invertible at $z=0$ and thus this is a well-defined coordinate transformation in each region with $\tau,z \in (-\infty, \infty)$. In these coordinates we can write the metric as
\begin{eqnarray}
    ds^2 = - ( 1-(R'(z))^2) d\tau^2 +\frac{R'(z)^2}{ 1-(R'(z))^2} dz^2 + R(z)^2 d \Omega^2\, .
\end{eqnarray}
\noindent Notice that by choosing $r = R(z)$ for $z>0$ or $z<0$ the metric has the form
\begin{eqnarray}
    ds^2 = - ( 1-A(r)^2) d\tau^2 +\frac{1}{ 1-A(r)^2}  dr^2 + r^2 d \Omega^2 \,.
\end{eqnarray}
However, we remark that the Jacobian of above transformation is $\pm 1$  for $z>0$ and $z<0$ respectively, which leads to different orientations. As a result, this requires us to define $\tau \to \pm \tau $ for $\pm z >0$.
\\

Note that each point in the $(t,r)$ coordinates can either correspond to $(\tau,z)$ or $(\tau,-z)$ as $R$ is quadratic in $z$. As a result, from
\begin{eqnarray}
    t = \tau-\tau_0(z) := \tau - \int_{z_0}^z dz' \frac{ (R')^2}{(R')^2 - 1}, \quad x = t+z =  \tau-\tau_0(z)  + z
\end{eqnarray}
we conclude that the two points in the $(t,x)$ coordinates are given by
\begin{eqnarray}
   (\tau-\tau_0(z), \tau-\tau_0(z) + z), \quad  (\tau+\tau_0(z), \tau+\tau_0(z) - z)\,.
\end{eqnarray}
We remark that 
\begin{eqnarray}
    \tau_0(z) - z = \int_{0}^z dz' \frac{ (R')^2}{(R')^2 - 1} - z = \int_{0}^z dz' \frac{ 1}{(R')^2 - 1} 
\end{eqnarray}
thus $\tau_0(z) - z \leq \tau_0(z) \leq 0 $ inside the inner horizon.
$(t,x)$ and $(t + 2\tau_0(x-t), -x + 2t+ 2\tau_0(x-t) $ are identified in figure \ref{fig:mapping-coord}.
~\\
Finally we want to note that the line given by $\hat{x}^{\mu}(t) := (t + 2\tau_0(x_s-t), -x_s + 2t+ 2\tau_0(x_s-t) $ for a given $x_s$ satisfies the geodesic equation
\begin{eqnarray}
    \frac{d^2\hat{x}^{\mu}}{dt^2} + \Gamma^{\mu}_{\rho \sigma}   \frac{d\hat{x}^{\rho}}{dt}\frac{d\hat{x}^{\sigma}}{dt} =0\,,
\end{eqnarray}
where $\Gamma^{\mu}_{\rho \sigma}$ are the Christoffel symbols.
\section{Canonical analysis of the mimetic model with \texorpdfstring{$\phi=t$}{}}
\label{app:cano_analysis}
\noindent 
In this section we will perform the canonical analysis of the mimetic model defined in \eqref{eq:cov_action} in the comoving gauge, i.e. the scalar field is set to $\phi=t$. As already studied in \cite{Han:2022rsx}, the gauge fixing of $\phi$ is exchangeable with the variation, thus we can directly work the gauge fixed action $S_2 |_{\phi = t, N =1}$.
In order to make touch with the effective LTB model in comoving gauge, the variables $X$ and $Y$ are set to
\begin{eqnarray}
    X = \frac{\partial_t E^{\phi} + \partial_x \left( N^x E^{\phi} \right)}{ E^{\phi}}  , \qquad Y = \frac{ \partial_t E^{x} + N^x \partial_x E^x}{ 2 E^{x}}\,.  
\end{eqnarray}
In total the corresponding Lagrangian up to boundary terms is given by
\begin{eqnarray}
    \mathcal{L}_2 &=&-\frac{3 {}_{(m)}\sqrt{1-4 \alpha^2 Y^2} E^{\phi} \sqrt{E^x}}{4 \alpha^2}+\frac{3 E^{\phi} \sqrt{E^x}}{4 \alpha^2}-\frac{ (X+Y) \sin^{-1}_{(m)}(2 \alpha Y) E^{\phi} \sqrt{E^x}}{2 \alpha} \\
&& -\frac{\partial_t E^{\phi} \partial_t E^x}{2 \sqrt{E^x}}+\frac{(\partial_x E^x)^2}{8 E^{\phi} \sqrt{E^x}}\nonumber +\frac{E^{\phi} (\partial_t E^x)^2}{8 (E^x)^{3/2}}+ \qty(X Y -\frac{Y^2}{2})  E^{\phi} \sqrt{E^x}+\frac{E^{\phi}}{2 \sqrt{E^x}}\,.\nonumber 
\end{eqnarray}
\noindent The generalized momenta of $E^{x}$ and $E^{\phi}$ are defined as
\begin{align}
    \pi_x &\coloneqq \frac{\partial \mathcal{L}_2}{ \partial 
(\partial_t E^{x})} = -\frac{\sqrt{E^x} \sin^{-1}_{(m)}\left(2 \alpha Y \right)}{2 \alpha} , \\
\pi_{\phi} &\coloneqq \frac{\partial \mathcal{L}_2}{ \partial 
(\partial_t E^{\phi})} = \frac{-E^{\phi} E^x }{4 \alpha (E^x)^{\frac{3}{2}} {}_{(m)}\sqrt{1-4 \alpha^2 Y}} \bigg[ 2 \alpha  \left( X- 2Y \right) +  {}_{(m)}\sqrt{1 -4\alpha^2 Y } \sin^{-1}_{(m)}\left(2 \alpha Y\right) \bigg] \,. \nonumber
\end{align}
Using this to perform the Legendre transformation of the Lagrangian we end up with
\begin{eqnarray}
    C^{\Delta}=-\frac{3 E^{\phi} \sqrt{E^x} \sin ^2\left(\frac{\alpha \pi_{\phi}} {\sqrt{E^x}}\right)}{2 \alpha^2}-\frac{\left(2 E^x \pi_x - E^{\phi} \pi_{\phi}\right) \sin \left(\frac{2 \alpha \pi_{\phi}} {\sqrt{E^x}}\right)}{2 \alpha}-\frac{(\partial_x E^x)^2}{8 E^{\phi} \sqrt{E^x}}-\frac{E^{\phi}}{2 \sqrt{E^x}}\,,
\end{eqnarray}
which reproduces exactly the Hamiltonian in \eqref{eq:effectiveHamiltonian} up to a boundary term after replacing the generalized momenta by extrinsic curvatures. The equations of motion of the triad variables are then given by
\begin{eqnarray}
    \partial_t E^x &=& - E^x \frac{\sin\left(\frac{2 \alpha \pi_{\phi} }{\sqrt{E^x}}\right)}{\alpha} + N^x \partial_x E^x, \\
    \partial_t E^{\phi} &=& \frac{\cos\left(\frac{2 \alpha \pi_{\phi} }{\sqrt{E^x}}\right) \left(E^{\phi} \pi_{\phi} - 2 E^x \pi_x \right)}{\sqrt{E^x}} - \frac{E^{\phi}\sin \left(\frac{2 \alpha \pi_{\phi} }{\sqrt{E^x}}\right)}{\alpha}\,.%
\end{eqnarray}

\section{Stationary solution of the generalized Einstein's equation \texorpdfstring{\eqref{eq:cov_eq}}{} in mimetic gravity}\label{app:cov_sol}
\noindent In order to derive the stationary solution of we start with the ansatz $h_{\nu\mu}(\tau,z) = h_{\nu\mu}(z)$ with $\nu,\mu \in \{\tau,z\}$ and further $\psi=\log(R(z))$. The mimetic condition the becomes
\begin{eqnarray}
   -h_{zz} (\partial_\tau \phi)^2 + (\partial_r \phi)^2 h_{\tau\tau} + h_{\tau\tau} h_{zz}=0\,,
\end{eqnarray}
which has the solution
\begin{eqnarray}
                       \phi = c_0 t + \int_{r_0}^R \phi'(r) dr, \qquad \phi'(r) = \sqrt{\frac{h_{zz}(1-h_{\tau\tau})}{h_{\tau\tau}}}(z)\,.
\end{eqnarray}
Plugging in the solution of $\phi$ in the variation $ \delta_{h_{11}}S=0$ we can determine the Lagrangian multiplier
\begin{eqnarray}
  \lambda = \frac{\left(\sqrt{ {\tilde{h}_{\tau\tau}}^2 }-1\right) \left(2 h_{\tau\tau} h_{zz} R''-R' \left(h_{zz} h_{\tau\tau}'+h_{\tau\tau} h_{zz}'\right)\right)}{2 h_{\tau\tau} h_{zz}^2 R \sqrt{{\tilde{h}_{\tau\tau}}^2}}\,,
\end{eqnarray}
where we defined
\begin{eqnarray}
    {\tilde{h}_{\tau\tau}}^2 := 1 - \frac{4 \alpha^2 (1-h_{\tau\tau}) R'^2}{h_{\tau\tau} h_{zz} R^2}\,.
\end{eqnarray}
Subsequently we can eliminate $\lambda$ in the remaining Lagrangian equations of motion $\delta_{h_{12}}S =\delta_{h_{22}}S = \delta_{\psi} S =0$ which are then equivalent to the two equations %
\begin{eqnarray}
   0=&&-\frac{h_{zz} \left({\tilde{h}_{\tau\tau}}^2-1\right) R^2-4 \alpha^2 R'^2}{32 \alpha^4 h_{zz}^2 \sqrt{{\tilde{h}_{\tau\tau}}^2} R'^2} \sqrt{\frac{\alpha^2 h_{zz} R'^2}{4 \alpha^2 R'^2-h_{zz} \left({\tilde{h}_{\tau\tau}}^2-1\right) R^2}}\\
   &&\Big(4 \alpha^2 R h_{zz}' R'-4 \alpha^2 h_{zz} \left(\sqrt{{\tilde{h}_{\tau\tau}}^2} R'^2+2 R R''\right) \nonumber \\
   &&+h_{zz}^2 \left(4 \alpha^2 \sqrt{{\tilde{h}_{\tau\tau}}^2}+\left(-4 {\tilde{h}_{\tau\tau}}^2+\sqrt{{\tilde{h}_{\tau\tau}}^2} \left({\tilde{h}_{\tau\tau}}^2+5\right)-2\right) R^2\right)\Big), \nonumber\\
   0=&&\frac{R'^2}{8 h_{zz} \sqrt{{\tilde{h}_{\tau\tau}}^2} \left(h_{zz} \left({\tilde{h}_{\tau\tau}}^2-1\right) R^2-4 \alpha^2 R'^2\right)^2 \sqrt{\frac{\alpha^2 h_{zz} R'^2}{4 \alpha^2 R'^2 -h_{zz} \left({\tilde{h}_{\tau\tau}}^2-1\right) R^2}}} \Big(-16 \alpha^4 \sqrt{{\tilde{h}_{\tau\tau}}^2} R'^4 \\
   &&-4 \alpha^2 \left({\tilde{h}_{\tau\tau}}^2-1\right) R^3 h_{zz}' R'+8 \alpha^2 h_{zz} \Big[\sqrt{{\tilde{h}_{\tau\tau}}^2} R'^2 \left(2 \alpha^2+\left({\tilde{h}_{\tau\tau}}^2+2\right) R^2\right) \nonumber\\
   &&-{\tilde{h}_{\tau\tau}} R^3 {\tilde{h}_{\tau\tau}'} R'+{\tilde{h}_{\tau\tau}}^2 R^2 \left(R R''-3 R'^2\right)-R^3 R''\Big] \nonumber\\
   &&-h_{zz}^2 R^2 \left(4 \alpha^2 \sqrt{{\tilde{h}_{\tau\tau}}^2} \left({\tilde{h}_{\tau\tau}}^2-1\right)+({\tilde{h}_{\tau\tau}}-1) ({\tilde{h}_{\tau\tau}}+1) \left(-4 {\tilde{h}_{\tau\tau}}^2+\sqrt{{\tilde{h}_{\tau\tau}}^2} \left({\tilde{h}_{\tau\tau}}^2+5\right)-2\right) R^2\right)\Big)\,. \nonumber
\end{eqnarray}
The solution of these equations can be expressed as
\begin{eqnarray}
     h_{\tau \tau} = 1 -  \frac{R_s}{R} +  \frac{\alpha^2 R_s^2}{R^4}, \qquad h_{zz} = \frac{c_0 (R')^2}{h_{\tau \tau}} , \ \quad R \geq \sqrt[3]{\alpha^2 R_s}
\end{eqnarray}
where $R_s$ is an integration constant. We can recover the form of metric in \eqref{metrictz1} by choosing $c_0 = 1$ and identifying $R_s = Gm_s$. 
Moreover, in the comoving gauge we have
\begin{eqnarray}
    X = \frac{\partial_t E^{\phi}}{ E^{\phi}} , \qquad Y = \frac{\partial_t E^{x}}{ 2 E^{x}} 
\end{eqnarray}
and at the bounce $\alpha b = \frac{\pi}{2}$, these variable approach the values
\begin{eqnarray}
    X \to -\frac{1}{z} + \frac{3 z}{2 \alpha^2}, \quad  Y \to -\frac{3 z}{2 \alpha^2}\,,
\end{eqnarray}
\begin{eqnarray}
    L = \mathcal{O}(z^2),\quad \partial_X L = \mathcal{O}(z^2),\quad \partial_X \partial_Y L = z^2 + \mathcal{O}(z^4) 
\end{eqnarray}
and thus the Lagrangian is still finite and well defined.

\section{Modified Junction condition}\label{app:junction}
\subsection{Junction condition in the mimetic model}
\noindent
We will now derive the polymerized junction condition in the framework of mimetic model with the Lagrangian given in \eqref{eq:cov_action}. We assume that the junction surface is embedded in the spacetime with tangent vector $\vec{l} = \partial_l$ and normal direction $\vec{n} = \partial_n $ in the 2D manifold ${\cal M}_2$, where the metric of the spherical space time can be expressed as
\begin{eqnarray}\label{eq:gaussian_coord_junction}
    ds^2 = \xi h_{nn}^2 dn^2 -\xi h_{ll}^2 dl^2 + h_{\theta \theta} d\Omega^2 = \xi d\tilde{n}^2 -\xi h_{ll}^2 dl^2 + h_{\theta \theta} d\Omega^2\,,
\end{eqnarray}
where $\xi = \pm 1 =\text{sign}(n^{\mu}n_{\mu})$ and $d \tilde{n} = |h_{nn}| dn =\text{sign}(\det e) h_{nn} dn$.
We introduce the 2D tetrad $e$ whose determinant is given by $\det e = h_{nn} h_{ll}$, such that the possible signature change of $\det e$ is captured.
We introduce some external sources which have the effective energy momentum tensor given by
\begin{eqnarray}
    T_{\mu \nu} = T_{\mu \nu}^{+} \Theta +  T_{\mu \nu}^{-} (1 - \Theta) + \tau_{\mu \nu} \delta(n) \,,
\end{eqnarray}
where we suppose the junction surface is determined by $n=0$. The mimetic density function is allow to have discontinuity as well where
\begin{eqnarray}
    \lambda =\lambda^{+} \Theta +  \lambda^{-} (1 - \Theta) + \delta \lambda \delta(n) \,.
\end{eqnarray}
The effective equation of motion in the corresponding mimetic model becomes
\begin{eqnarray}
     \tilde{G}_{\mu\nu} = G_{\mu\nu}- T^{\phi}_{\mu\nu} =\text{sign}(\det e) \left( \frac{\kappa}{2} T_{\mu\nu} - 2\lambda\partial_{\mu} \phi\partial_{\nu} \phi \right)\, ,
\end{eqnarray}
which is a modified Einstein equation in the distributional sense. Note here we use the coupling of external matter sources $T_{\mu\nu}$ and the Mimetic energy density $\lambda$ with parity even coupling $\sqrt{-g}$ instead of $\det e$, otherwise a $\text{sign}(\det e)$ will appear in front of $\sigma_{\mu\nu}$ and $\delta \lambda$.  From \eqref{eq:gaussian_coord_junction}
we define the extrinsic curvature adapting to the appearance of $\det e$:
\begin{eqnarray}
    K_{ll} = \xi \frac{ h_{ll} \partial_n h_{ll}}{ h_{nn} }, \qquad  K_{\theta \theta} =  \xi \frac{\partial_n h_{\theta\theta}}{ 2 h_{nn} } \ .
\end{eqnarray}
The above equation implies that, when we have a timelike junction surface $\xi=1$ and possible signature changes of $\det e$, i.e. $h_{nn}$ may take different signatures, we will take the normalized normal vector pointing increasing $n$ direction to be $\partial_n/h_{nn} $ instead of $\partial_n/|h_{nn}| $. Equivalently, this implies that for a normalized normal vector $\tilde{n}$ we have
\begin{eqnarray}
    K_{ij} = \text{sign}(\det e) \frac{\partial_{\tilde{n}} h_{ij}}{2} \,,
\end{eqnarray}
where we use $d \tilde{n} = |h_{nn}| dn =\text{sign}(\det e) h_{nn} dn$.
The distributional tensorial
fields are supposed to be appearing in second derivatives along the $n$ direction, e.g. $\partial_n K_{ij}$ terms and $\partial_{n}^2 \phi$ terms. For possibly non-smooth $g = g_1(n)\Theta(n) + g_2(n)(1-\Theta(n))$ we define
\begin{eqnarray}                                                                   
  \lim_{\epsilon \to 0} \int_{-\epsilon}^{\epsilon}dn f(n) \partial_n g_n = \lim_{\epsilon \to 0} \int_{-\epsilon}^{\epsilon}dn f(n) \left( g_1'(n)\Theta(n) + g_2'(n)(1-\Theta(n)) +[g(n)]\delta(n) \right)= [g(0)]\overline{f}(0) 
  \ , \nonumber \\
\end{eqnarray}
where we define $\overline{f}(n) := (f_1(n)+f_2(n))/2$ to allow non-smooth functions $f$.
~\\
Notice that from the mimetic condition $\partial_{\mu} \phi \partial^{\mu} \phi = -1$ we can transform derivatives  of $\phi$ along $n$ to derivatives of $\phi$ along $l$ and derivatives of metric derivatives along $n$, namely
\begin{eqnarray}
    \partial_n \phi = s_{\phi} \frac{h_{nn} \sqrt{(\partial_l \phi)^2 - \xi h_{ll}^2}}{h_{ll}} , \qquad s_{\phi} = \pm 1.
\end{eqnarray}
Thus we have
\begin{eqnarray}
    -2 \lambda \partial_{n} \phi \partial_{n} \phi =2h_{nn}^2 \left(1-\xi \frac{\partial_l \phi^2}{h_{ll}^2}\right), \quad -2 \lambda \partial_{n} \phi \partial_{l} \phi = - s_{\phi} \frac{2h_{nn} \partial_l \phi \sqrt{\partial_l \phi^2-\xi h_{ll}^2}}{h_{ll}} .
\end{eqnarray}
As a result, all distributional contribution from derivatives of $\phi$ will be encoded in $\partial_n K_{ij}$ as we assume continuous derivatives along the junction surface for $\partial_l \phi$ and $\partial_l^2 \phi$.
Now plugging in the ansatz of our $L$ function, we obtain that the distributional terms in the modified Einstein tensor reads %
\begin{eqnarray*}
    \tilde{G}_{n}{}^{l} &=&s_{\phi} \xi \frac{\partial_l \phi \sqrt{h_{ll}^2 - \xi \partial_l \phi^2} \left( ( (-1)^m - \sqrt{1-4 \alpha^2 Y^2})  %
    \right)}{ h_{ll}^3 h_{\theta\theta} \sqrt{1-4 \alpha^2 Y^2}} \partial_n K_{\theta\theta}\\ %
    \tilde{G}_{n}{}^{n} &=& - \xi \frac{\left(h_{ll}^2- \xi \partial_l \phi^2\right) \left({ \left({(-1)^m}{\sqrt{1-4 \alpha^2 Y^2}}-1\right)}
    \right)}{{h_{\theta\theta}} h_{nn} h_{ll}^2} \partial_n K_{\theta\theta}\\ %
    \tilde{G}_{l}{}^{l} &=&\frac{ \left(\partial_l \phi^2 \left(\sqrt{1-4 \alpha^2 Y^2}-(-1)^m\right) + \xi (-1)^m h_{ll}^2\right)}{h_{nn} h_{ll}^2 h_{\theta\theta} \sqrt{1-4 \alpha^2 Y^2}} \partial_n K_{\theta\theta} %
    \\ %
    \tilde{G}_{\theta}{}^{\theta} &=&\partial_n K_{\theta\theta} \frac{{ \left(\xi (-1)^m \left(h_{ll}^2+ \xi \partial_l \phi^2\right) \left(4  \alpha^2 Y (X-3 Y)+1\right)+\partial_l \phi^2 \left(1 -4 \alpha^2 Y^2\right) \sqrt{1-4 \alpha^2 Y^2}\right)}}{2 h_{nn} h_{ll}^2 h_{\theta \theta} \left(1-4 \alpha^2 Y^2\right)^{3/2}} \nonumber \\
    && + \partial_n K_{ll}\frac{(-1)^m  \left(4 \alpha^2 Y^2-1\right)}{h_{nn} h_{ll}^2 \left(1-4 \alpha^2 Y^2\right)^{3/2}} .
\end{eqnarray*}

\noindent
From $\lim_{\epsilon \to 0} \int_{-\epsilon}^{\epsilon}d \tilde{n} \tilde{G}_{\mu \nu} =\lim_{\epsilon \to 0} \int_{-\epsilon}^{\epsilon}d \tilde{n} \left( \frac{\kappa}{2} T_{\mu\nu} -2 \lambda \partial_{\mu} \phi \partial_{\nu} \phi \right) $, we obtain in general for $\xi=1$, namely the junction surface is timelike,
\begin{align}
[ K_{\theta}{}^{\theta} ] &= -\frac{\kappa}{4} (-1)^m \sqrt{1-4 \alpha^2 Y^2} \sigma_{l}{}^{l}\\
\delta \lambda &= \frac{\kappa}{8} \sigma_{l}{}^{l} \left(1-(-1)^m \sqrt{1-4 \alpha^2 Y^2} \right)\\
[K_{l}{}^{l}] &= - \frac{\kappa}{4  (-1)^m \sqrt{1-4 \alpha^2 Y^2} } \Big[\left(-4 \alpha^2 Y ((\sigma_{\theta}^{\theta}-3 \sigma_{l}^{l}) Y+\sigma_{l}^{l} X)-\sigma_{l}^{l}+2\sigma_{\theta}^{\theta}\right) \\
&\quad-\sigma_{l}^{l} (\partial_l \phi)^2 \Big(4 \alpha^2 Y \left(-\xi \left(X - 3Y \right)+(-1)^m Y \sqrt{1-4 \alpha^2 Y^2}\right) -(-1)^m \sqrt{1-4 \alpha^2 Y^2}+1\Big)\Big] \,  . \nonumber
\end{align}
There is a special case when $h_{ll}^2 = \xi (\partial_l \phi)^2$, or equivalently $\partial_n \phi =0$, in which case the condition becomes
\begin{align}
    [ K_{\theta}{}^{\theta} ]  &= - \frac{\kappa}{4} \left( \sigma_{l}^{l}-2 \delta \lambda \right)\\
   [K_{l}{}^{l}] &= \frac{\kappa}{4} (-1)^{m} \sqrt{1-4 \alpha^2 Y^2} (-2 \delta \lambda+\sigma_{l}^{l}-2\sigma_{\theta}^{\theta})
\end{align}
and $\delta \lambda$ is completely free.

\subsection{Junction condition for the OS collapse}\label{app:junction_noshock}
For the dust mass shell we have the metric
\begin{eqnarray}
    ds^2 = - dt^2 + x^2 a(t)^2 (dx^2 + d \Omega^2)
\end{eqnarray}
thus on a $x=x_s =const$ surface we have
\begin{eqnarray}
    K_{t}^{t} = 0, \quad K_{\phi}^{\phi} = K_{\theta }^{\theta} = \frac{1}{x_s a(\tau)}
\end{eqnarray}
with 
\begin{eqnarray}
    a(t) = \left( 2G E^0 \left( \frac{9}{4}  (-t)^2+ 
\alpha^2 \right)\right)^{\frac{2}{3}}\,.
\end{eqnarray}
Moreover, for the dust mass shell we have $\partial_x \phi =0$, which is also satisfied for the exterior vacuum metric.
On the other hand, for the exterior vacuum metric in the $(\tau,z)$ coordinates, we have
\begin{eqnarray}
    ds^2 = -(1-R'(z)^2) d \tau^2 + \frac{R'(z)^2}{1-R'(z)^2} dz^2 +  R(z )^2 d \Omega^2 ,
\end{eqnarray}
The continuity of the metric on the junction surface implies $R(z(t)) = x_s a(t)$ thus $z(t) = \pm \left( x_s - t \right)$ and $\tau'(t)= \pm \frac{1}{1-R'(z(t))^2}$. One can check that this is compatible with $\tau = x_s + s_1 (\tau_0(z)-z), s_1 = \pm 1$ and thus implies that the junction surface is determined by $x=x_s$ and its $z \to -z$ counter part in $(t,x)$ coordinates. We then calculate the tangent and normal vector along the line which is given by
\begin{eqnarray}\label{eq:normal_junc_os}
  \vec{l}= \pm \left( \frac{s_1}{1-R'(z)^2},-1 \right),\quad  \vec{n}= s_2 \left( \frac{ R'(z)}{1-R'(z)^2},-\frac{s_1}{R'(z)}\right), 
\end{eqnarray}
One can then further check that
\begin{eqnarray}\label{eq:extrinsic_vacuum_z}
    K_{l}{}^{l}= 0, \qquad K_{\phi}{}^{\phi} = K_{\theta }{}^{\theta} = \frac{1}{x_s a(\tau)} = - s_1 s_2 \ \text{sign}(R'(z)) \frac{1 }{ R} \, .
\end{eqnarray}
Note that since for $s_1 = \pm 1$ before the bounce we have $R'(z) = \pm 1$ and after the bounce $R'(z) = \mp 1$. As can be seen from figure \ref{fig:mapping-coord} and \eqref{eq:normal_junc_os}, 
for $z>0$ part we have $s_2 = -s_1$ in order to have a normal vector pointing to the interior of the vacuum region, i.e. the increasing $z$ direction. For $z<0$ and $s1=1$ we have $R'(z)<0$ thus $s_2 = 1$. Thus from \eqref{eq:extrinsic_vacuum_z} we see for both case we have a same sign of $K_{\theta}{}^{\theta}$ thus $[K_{\theta}{}^{\theta}] = 0$

Now we need to check the continuity of the derivatives of $\phi$ function across the junction condition. From Appendix. \ref{app:cov_sol} we see that in $(\tau, z)$ coordinate
\begin{eqnarray}
    \partial_t \phi= s_{\phi}^{(1)} :=\pm 1, \qquad \partial_{x} \phi = s_{\phi}^{(2)} \frac{R'(z)^2}{1-R'(z)^2} , \quad s_{\phi}^{(2)} =\pm 1
\end{eqnarray}
This implies on the junction surface
\begin{eqnarray}
    \partial_n \phi = \vec{n} \cdot \nabla \phi = s_2 ( s_{\phi}^{(2)} s_1 + s_{\phi}^{(1)})\frac{R'(z)}{(1-R'(z)^2)} , \quad \partial_l \phi = \vec{l}  \cdot \nabla \phi = \frac{s_1 s_{\phi}^{(1)} +    s_{\phi}^{(2)}  R'(z)^2}{(1-R'(z)^2)} .
\end{eqnarray}
Since for the homogeneous collapse we have $\partial_n \phi = 0$, this implies, when $s_1$ changes sign, the sign difference between $ s_{\phi}^{(1)} = - s_1 s_{\phi}^{(2)} $ must change as well. For the case where $s_1 =1$ for both before or after the bounce, we can set $ s_{\phi}^{(1)}  = 1$ such that $\partial_l \phi = 1$ along the junction surface for the exterior vacuum solution as well. In this case we have a consistent solution of $\phi$ for the whole spacetime. However, when we have different signs of $s_1$ before and after the bounce, i.e. $s_1 =1$ before and $s_1 = -1$ after the bounce where the vacuum solution corresponds to the solution of $z>0$ only, we can not have a continuous  %
$\partial_l \phi^2$ across the junction condition.

We remark that the junction surface for $\pm z$ can be check in the $(t,x)$ coordinates as well. This corresponds to explicitly the coordinate transformation from $(\tau,\pm z)$ to $(t,x)$.

\subsection{No shock solution without external matter field for a continuous mimetic field}\label{app:junction_shock}
Now we will explore if we can have a shock solution purely produced by a discontinuity in $\delta \lambda$ without introducing external fields. In this analysis we restrict our analysis to a continuous $\partial_n \phi$ field across the junction surface. According to the junction condition \eqref{eq:junction_condition_general}, the vanishing of the external field with a non-trivial $\delta \lambda$ is only possible when we have $\partial_n \phi = 0$ on the junction surface, and we use \eqref{eq:junction_condition_dp0}. This is purely a constraint coming from the mimetic part which yields the restriction that on the orthogonal direction we have $\partial_l \phi \partial^l \phi = 1$. Thus after gauge fixing we have a Hamiltonian in the form of \eqref{eq:effectiveHamiltonian} with $l$ being the temporal direction. Since we have in the $(t,x)$ coordinate $\partial_t \phi = 1$ and $\partial_{x} \phi = 0$, on any surface determined by $(t(t_j),x(t_j))$ where $n=\frac{x'(t_j) R'(z)}{\sqrt{t'(t_j)^2 - x'(t_j)^2 R'(z)^2}}\left( 1, \frac{t'(t_j)}{x'(t_j) R'(z)^2}\right)$ labels the orthogonal direction, we have
\begin{eqnarray}
    \partial_{t_j} \phi &=& t'(t_j) \partial_t \phi + x'(t_j) \partial_{x} \phi = t'(t_j) \partial_t \phi \, , \\
    \partial_n \phi &=& \frac{x'(t_j) R'(z)}{\sqrt{t'(t_j)^2 - x'(t_j)^2 R'(z)^2}} \partial_t \phi + \frac{t'(t_j)}{\sqrt{t'(t_j)^2 - x'(t_j)^2 R'(z)^2}R'(z)} \partial_{x} \phi
\end{eqnarray}
As a result, for $t_j$ to be an affine parametrization, which requires $\partial_{t_j} \phi  =1, \partial_n \phi = 0$, we have $t_j=t$ and $x'(t_j) = 0$. However, this is nothing else but the junction surface studied in the previous section which has $\delta \lambda = 0$. 

Since the junction surface is timelike, such result coincide with the initial value problem imposed by the effective model.

\end{document}